\documentclass[12pt]{article}
\usepackage{color}
\newcommand{\beq}{\begin{equation}}
\newcommand{\eeq}{\end{equation}}
\newcommand{\beqa}{\begin{eqnarray}}
\newcommand{\eeqa}{\end{eqnarray}}
\newcommand{\beqar}{\begin{eqnarray*}}
\newcommand{\eeqar}{\end{eqnarray*}}

\newcommand{\al}{\alpha}
\newcommand{\be}{\beta}

\def\spa          {\ \ \ }
\def\non          {\nonumber}
\def\ha           {\mbox{$\frac{1}{2}$}}

\def\spa          {\ \ \ }
\def\mand         {\spa\mbox{and}\spa}

\def\Tr           {\mbox{\rm Tr}\,}

\def\cd           {{\cdot}}
\def\ran          {\rangle}
\def\lan          {\langle}

\def\fsH    {H\!\!\!\!/\,}
\def\fsC    {C\!\!\!\!/\,}
\newcommand{\del}{\delta}

\newcommand{\eps}{\epsilon}
\newcommand{\ga}{\gamma}
\newcommand{\Ga}{\Gamma}

\newcommand{\inn}{\!\cdot\!}

\newcommand{\lam}{\lambda}

\newcommand{\sig}{\sigma}

\newcommand{\labell}[1]{\label{#1}} 
\newcommand{\reef}[1]{(\ref{#1})}

\def\sst#1{{\scriptscriptstyle #1}}
\def\0{{\sst{(0)}}}
\def\1{{\sst{(1)}}}
\def\2{{\sst{(2)}}}
\def\3{{\sst{(3)}}}
\def\4{{\sst{(4)}}}
\def\5{{\sst{(5)}}}
\def\6{{\sst{(6)}}}
\def\7{{\sst{(7)}}}
\def\8{{\sst{(8)}}}

\begin{document}
\baselineskip 18pt%
\begin{titlepage}
\vspace*{1mm}%
\hfill
\vbox{

    \halign{#\hfil         \cr
           } 
      }  
\vspace*{9mm}
\vspace*{9mm}%

\center{ {\bf \Large
Remarks on the mixed Ramond -Ramond, open string scattering amplitudes of BPS, non-BPS and brane-anti brane
}}\vspace*{1mm} \centerline{{\Large {\bf  }}}

\begin{center}
{Ehsan Hatefi   \small $^{1,2,3}$}

\vspace*{0.3cm}{ {\it 
\vskip.1in
{ $^{1}$ Centre for Research in String Theory, School of Physics and Astronomy,
Queen Mary University of London,Mile End Road, London E1 4NS, UK},
\vskip.06in
{ $^{2}$National Institute for Theoretical Physics , School of Physics and Mandelstam Institute for
Theoretical Physics,University of the Witwatersrand, Wits, 2050, SA},
\vskip.06in
{ $^{3}$ Institute des Hautes Etudes Scientifiques Bures-sur-Yvette, F-91440, France }


}}
\vspace*{0.1cm}
\vspace*{.1cm}
\end{center}
\begin{center}{\bf Abstract}\end{center}
\begin{quote}

From the world-sheet point of view we  compute  three, four and five point BPS and  non-BPS scattering amplitudes of type IIA and IIB superstring theory.
All these the mixed S-matrix elements including a closed string Ramond-Ramond (RR) in the bulk and a scalar/gauge or tachyons with  their all different pictures ( including RR in asymmetric and symmetric pictures)  have been carried out.
We have also shown that in  asymmetric pictures various equations must be kept fixed. More importantly, by direct calculations on upper half plane, it is realized that some of the equations (that must be true) for BPS branes can not be necessarily applied to non-BPS amplitudes.
We also derive the S-Matrix elements of $<V_C^{-2} V_{\phi}^{0}V _A^{0} V_T^{0}  >$  and clarify the fact that in the presence of the scalar field and RR, the terms carrying momentum of RR in transverse directions  play important role in the entire form of the S-matrix and their presence is needed in order to have gauge invariance for the entire S-matrix elements of type IIA (IIB) superstring theory.

\end{quote}
\end{titlepage}

\section{Introduction}

In a very important paper \cite{Polchinski:1995mt}, it has been extensively clarified that the sources for all different kinds of D-branes  are  Ramond- Ramond (RR) fields. It is worthwhile looking at some concrete references regarding RR fields  \cite{Witten:1995im,Polchinski:1996na}. Besides them, RR fields play a very crucial effect in understanding the phenomenon of Dielectric branes which was first demonstrated by Myers  in \cite{Myers:1999ps}.
Having employed several RR couplings of \cite{Hatefi:2012sy}, we  could explore and interpret the $N^3$ entropy of $M5$ branes as well.    
\vskip.1in

 It is also known that if one wants to work with the effective actions of type IIA,IIB string theory, then one needs to deal with DBI and Chern-Simons effective actions which can be accordingly found in  \cite{Bachas:1995kx},\cite{Li:1995pq},\cite{Douglas:1995bn} and \cite{Green:1996dd}\footnote {Some of the new curvature corrections of type II have been recently obtained in \cite{Junghans:2014zla}.}.

\vskip.1in

By making use of the scattering theory of D-branes in the world volume of BPS branes in type II string theory we have also explored various new Chern-Simons couplings  including their all order $\alpha'$ corrections to the low energy effective actions of D-branes. In fact it is in detail explained that for BPS (non-BPS) branes   all the  corrections to D-brane effective actions can just be derived by having the entire form of S-matrix  and not by any other tools like T-duality transformation. The reason  for this sharp conclusion  is that by having S-matrix we are able to actually fix all the coefficients of the effective field theory couplings (and also their higher order $\alpha'$ corrections)  without any ambiguities ( see  \cite{Hatefi:2012zh,Hatefi:2010ik} for BPS  and \cite{Hatefi:2012wj} for non-BPS branes). In fact all the three different ways of obtaining the couplings in the effective field theory involving pull-back, Taylor expansions and new Wess-Zumino terms  (the generalization to Myers action)
have also been explained in \cite{Hatefi:2012zh}.

\vskip .2in

It is emphasized  in \cite{Sen:1999mg,Lambert:2003zr,Sen:2004nf} that to get to the effective theory of non supersymmetric cases or non-BPS branes one has to integrate out all the massive modes and needs to effectively work out  just with the massless strings such as scalar, gauge fields and also employ the real components of  tachyon fields.
 \vskip .1in

There are various motivations to perform scattering theory of all BPS and non-BPS branes , basically one of the main reasons to employ it, is to actually have the entire form of   S-matrix elements   which is a physical quantity and the other motivation would be to deal with its strong potential of  gaining new terms including  their corrections (of course  without any ambiguity) to string theory effective actions.

Having set this formalism, one may end up obtaining several new couplings for all BPS branes in RNS formalism \cite{Hatefi:2013eia,Barreiro:2013dpa} and eventually one could investigate to get to a proposal for the corrections to some of the couplings  \cite{Hatefi:2012rx}. In particular  it is shown that some of these couplings must be employed to be able to  work with some of the applications of either M-theory \cite{Maxfield:2013wka} or gauge-gravity duality \cite{Hatefi:2012bp}.

\vskip .2in

One might  look for  various applications in the world volume  of non-BPS branes as some of them were comprehensively pointed out in  \cite{Hatefi:2012wj}, but for the concreteness we highlight some of them again.

As an instance in  \cite{deAlwis:2013gka} it has been illustrated that  as long as the effective field theory description holds, in the large volume case (despite having the non supersymmetric cases) , the Ads minima  are indeed stable vacua.
Not only  the phenomena of the production of branes are investigated in     \cite{Bergman:1998xv,Witten:1998cd}  but also  Inflation in the language of D-brane and anti D-brane systems in string theory was revealed \cite{Dvali:1998pa,Burgess:2001fx,Choudhury:2003vr,Kachru:2003sx}. It is also worth mentioning that tachyons of type IIA(IIB) superstring theory (with odd-parity) have been taken into account to make various remarks on some holographic QCD models  \cite{Casero:2007ae,Dhar:2007bz}
as well.

\vskip .3in

 In \cite{Friedan:1985ge} various issues on the scattering amplitudes have been extensively discussed, however, one has to concern about some other issues on the mixed amplitudes involving a closed string Ramond-Ramond (RR) and some other open strings  for which some of them have been empirically addressed in \cite{Hatefi:2012rx}. The content of this paper
is beyond what has been appeared in those references. Indeed we would like to understand in the presence of  symmetric and asymmetric picture of RR, a scalar field and some other open strings what happens to the terms that carrying momentum of RR in transverse directions, arguing that the S-matrices that satisfy Ward identity and do involve all the contact interactions are definitely the correct S-matrices.

\vskip.2in

The paper organized as follows. In the next section, first we try to come up with the entire details of a three point  \footnote{three point function from the world sheet point of view and  two point function from the space time prospective} scattering computations of a closed string  RR in asymmetric picture $(C^{-2}$ picture, in terms of its potential not its field strength) and a scalar field. Then we carry out the same S-matrix in symmetric picture of RR  $(C^{-1}$ picture, in terms of RR's field strength) and scalar picture in $(-1)$ picture. Finally we compare both S-matrixes and make remarks on a Bianchi identity that must be true to get to picture independence result. For the completeness we  talk about $CT,CA$ amplitudes as well.

\vskip.1in

The four point correlation function of $<T^{(0)}T^{(0)} C^{(-2)}>$ in the world volume of D-brane anti D-brane system has also been carried out to show that the result is the same as $<T^{(-1)}T^{(0)} C^{(-1)}>$ \cite{Kennedy:1999nn}. This clearly confirms that there is no any issue of picture dependent of the mixed closed string RR and strings that move on the world volume space  such as gauge fields or tachyons (but not scalar field).\footnote{ Because the polarization of scalar field is in the bulk and there is a non zero correlator between RR and the first term of scalar vertex operator in zero picture as $<e^{ip.x(z)} \partial_i x^i(x_1)> $ is non zero. Basically one needs to concern about the terms that carry momentum of RR in transverse directions ($ p^{i}$'s terms).}
\vskip.2in

Hence due to momentum  conservation along the world volume of branes and as long as we are dealing with world volume gauge fields and or  tachyons in the presence of RR, there is no any issue about choosing the picture of RR (in symmetric or asymmetric ) \footnote{In fact in order to get to the final answer for these S-matrices as fast as it is possible, one could put RR and a gauge field in $(-1)$ picture  and the other tachyons/gauges in zero picture.}.

\vskip.1in

However, because there is a non-zero correlation function between RR and scalar in zero picture and also due to the fact that the terms carrying momentum of RR in transverse directions $(p^i,p^j)$ can not be derived by any duality transformations \cite{Hatefi:2012zh,Hatefi:2012wj,Hatefi:2012ve},  one has to be concerned about various issues. In the following we release  various subtleties that have to be considered for the mixed amplitudes involving a closed string RR, a scalar and some arbitrary numbers of gauges or tachyons in the world volume of  BPS, brane/anti brane  and non-BPS systems.

\vskip.1in

We then start calculating  all the four point non-BPS functions of type II in their all different pictures   $<T^{(-1)}\phi^{(0)} C^{(-1)}>$ ,$<T^{(0)}\phi^{(-1)} C^{(-1)}>$,
$<T^{(0)}\phi^{(0)} C^{(-2)}>$. In the case of $<T^{(-1)}\phi^{(0)} C^{(-1)}>$ we  see  the term that carries momentum of RR in transverse direction disappears after applying a Bianchi identity equation, however, we claim that one has to be careful about these transverse $(p^i,p^j)$ terms in higher point functions as their presence plays crucial role in the gauge invariance of the higher point mixed amplitudes. 
There is a non-zero correlation function between RR  and the first part of the vertex operator of scalar field in zero picture and therefore one needs to think  about those terms that carry momentum of RR in transverse direction ($p^i,p^j$ ) as they can not be derived by any duality transformation \cite{Hatefi:2012zh}. Indeed by direct computations of scattering amplitudes of BPS branes, we observe that several Bianchi identity must hold for BPS cases, whereas we show that these equations can not be  necessarily true for non-BPS branes (say for $<T^{(0)}T^{(0)}T^{(0)} C^{(-2)}>$) ,otherwise the whole   non supersymmetric S-matrix vanishes. For the completeness we perform $<\phi^{(0)}A^{(0)} C^{(-2)}>$, $<\phi^{(-1)}A^{(0)} C^{(-1)}>$ and $<\phi^{(0)}A^{(-1)} C^{(-1)}>$ as well.

\vskip.1in

We also would like to go over to some of the mixed RR, scalar  and tachyon five point functions of either brane/ anti brane or non-BPS branes to see what happens, if we carry them out in both symmetric and asymmetric picture of RR  accordingly. In fact for a scattering amplitude of brane anti brane system (including scalar, tachyons and RR), we have three different choices.\footnote{ $<\phi^{(-1)} T^{(0)} T^{(0)} C^{(-1)}>,<\phi^{(0)} T^{(-1)} T^{(0)} C^{(-1)}>$ and  $<\phi^{(0)} T^{(0)} T^{(0)} C^{(-2)}>$
.} Indeed for these particular amplitudes there is no Ward identity and a priori one does not know which specific picture gives us the correct S-matrix, where we claim and establish the fact that already at the level of S-matrix for brane / anti brane, one needs to know some generalised  Bianchi identities to be able to produce all the effective field theory couplings.

 Most importantly, we show that  the terms that carry momentum of RR in transverse directions ($p^i,p^j$), that are singular  of $<\phi^{(0)} T^{(-1)} T^{(0)} C^{(-1)}>$ and  $<\phi^{(0)} T^{(0)} T^{(0)}C^{(-2)}>$ amplitudes remain after taking integrations properly on the upper half plane, while these terms are absent in $<\phi^{(-1)} T^{(0)} T^{(0)} C^{(-1)}>$.  Hence  in order to remove all the apparent singularities of brane/anti brane, we introduce new Bianchi identities at the level of world-sheet. 
 
 \vskip.2in
 
  We then move on to  obtain   $<T^{(0)} T^{(0)} T^{(0)} C^{(-2)}>$ S-matrix and 
  determine the fact  that if we apply some of the Bianchi identities of BPS branes to this non-BPS amplitude  then the whole S-matrix disappears so this clearly confirms  that those Bianchi equations of BPS branes must not be true  for non-BPS amplitudes. The reason is that, there are non zero field theory couplings in the world volume of non-BPS branes that have to be produced by non-zero S-matrix of an RR and three tachyons.  We will also mention several subtleties that might have potentially something to do with some of the issues on the perturbative string theory that are released and have been pointed out in a series of papers appeared by Witten \cite{Witten:2012bh}.

\vskip .1in
The five point correlations of $ <V_C^{-1} V_{\phi}^{-1} V_T^{0} V_A^{0}>$ have been computed in \cite{Hatefi:2013yxa}. In order to get to know what happens to the gauge invariance of the amplitudes we come over to the same amplitude but with different picture of scalar as   $<V_C^{-1} V_{\phi}^{0} V_A^{-1} V_T^{0} >$ .

  Given the fact that the vertex operator of a scalar field in zero picture carries two different terms and in particular its first part has non zero correlation function with closed string RR, we see that here the terms carrying momentum of RR (all $p.\xi$ terms) survive after taking integration on upper half plane.  In fact due to all non vanishing $p.\xi$ terms the final form of the S-matrix does not satisfy Ward identity associated to gauge field unless we introduce new Bianchi identities.  Since we can not give up  gauge  invariance of the S-matrix, we need to  come up with some ideas. That is why we  look at the same S-matrix in an asymmetric picture of RR $<V_C^{-2} V_{\phi}^{0} V_T^{0} V_A^{0}>$ where in this picture  upon considering  the known Bianchi identities not only  will we observe that the amplitude respects the Ward identity associated to the gauge field, but also we are able to get to the whole contact interactions of the related S-matrix.

  \vskip.1in

   We may wonder why we can not see the term that carries momentum of RR in transverse direction in four  point functions, say $ <V_C^{-1} V_{\phi}^{0} V_A^{-1}>$. The answer is that in these functions we do have that particular  term , however, after gauge fixing the integral should be taken on the whole space time (from $-\infty$ to $\infty$) and since the integrand ( including $p^i$ term) is odd , the final result naturally is zero. But this does not happen for five point functions any more, basically, for five and higher  point functions after fixing SL(2,R) invariance we need to take the integrals on the position of closed string and the remaining terms  involving ($p^i,p^j$) terms are not vanished. Even these terms might not satisfy Ward identity. The resolution to this,  is to either introduce some new Bianchi identities or 
 calculate all of the mixed S-matrixes in symmetric/ asymmetric pictures. Let us get to the technical parts by computing three point functions.

\vskip .2in


\section{The $\phi^{(0)}-C^{(-2)}$ amplitude}

In this section we are going to derive the full S-matrix elements of one scalar field and one closed string Ramond-Ramond (RR) in type IIA (IIB) String theory, where  for some reasons we would like to keep RR in its asymmetric picture. That is, we consider   its vertex operator in terms of its potential ( not its field strength) so we deal with RR in  $C^{-3/2,-1/2}$ picture.  
 \vskip.1in

 Motivation for doing this computation in different pictures is that , there is no  Ward identity for the mixed RR and a scalar field and one might wonder what happens  in higher point functions of all mixed amplitudes including RR, scalar field and tachyons (but not gauge field), or one might ask which particular picture is going to give us the complete S-matrix elements including all the infinite contact interactions of string theory.

Note that our notations are  such that $\mu,\nu,..$ run over the whole space-time, $a,b,c,..$ and $i,j,k,..$  are world volume and transverse directions accordingly.
 
 Thus, this  four point function from world-sheet  point of view  ( three point function, from the  space-time view) of one scalar and an asymmetric  RR closed string is given by its following correlation function 
\beqa
{\cal A}^ {C^{(-2)}\phi^{(0)}} & \sim & \int dxdz d\bar z
 \lan V_{RR}^{(-2)}(z,\bar{z})   V_\phi^{(0)}(x)\ran\labell{coraa}\eeqa
Note that  the vertex of RR in asymmetric picture has been first proposed by \cite{Bianchi:1991eu} so the vertices can be read off as follows
\beqa
V_{\phi}^{(0)}(x) &=& \xi_{1i}(\partial^i X(x)+i\alpha'k.\psi\psi^i(x))e^{\alpha'ik.X(x)}\nonumber\\
V_{RR}^{(-2)}(z,\bar{z})&=&(P_{-}\fsC_{(n-1)}M_p)^{\al\be}e^{-3\phi(z)/2} S_{\al}(z)e^{i\frac{\alpha'}{2}p\cd X(z)}e^{-\phi(\bar{z})/2} S_{\be}(\bar{z}) e^{i\frac{\alpha'}{2}p\cd D \cd X(\bar{z})}\nonumber
\eeqa

We are considering the disk level amplitude so the RR has to be put in the middle of disk while the scalar field needs to be replaced just in its boundary. On-shell conditions for scalar and RR are \footnote { The definition of projector and the field strength of closed string is

\begin{displaymath}
P_{-} =\ha (1-\ga^{11}), \quad
\fsH_{(n)} = \frac{a
_n}{n!}H_{\mu_{1}\ldots\mu_{n}}\ga^{\mu_{1}}\ldots
\ga^{\mu_{n}}
\non\end{displaymath} where  for type IIA  (type IIB) $n=2,4$,$a_n=i$  ($n=1,3,5$,$a_n=1$) with   $(P_{-}\fsH_{(n)})^{\al\be} =
C^{\al\del}(P_{-}\fsH_{(n)})_{\del}{}^{\be}$

 notation for spinor .}
\beqa
  k^2=p^2=0, \quad k.\xi_1=0
\nonumber\eeqa

For simplicity it is really useful to just make use of World-sheet's holomorphic elements, which means that we employ some valuable change of variables as follows

\begin{displaymath}
\tilde{X}^{\mu}(\bar{z}) \rightarrow D^{\mu}_{\nu}X^{\nu}(\bar{z}) \ ,
\spa
\tilde{\psi}^{\mu}(\bar{z}) \rightarrow
D^{\mu}_{\nu}\psi^{\nu}(\bar{z}) \ ,
\spa
\tilde{\phi}(\bar{z}) \rightarrow \phi(\bar{z})\,, \mand
\tilde{S}_{\al}(\bar{z}) \rightarrow M_{\al}{}^{\be}{S}_{\be}(\bar{z})
 \ ,
\non\end{displaymath}

Some definitions might be  important to highlight as well . \footnote{
\begin{displaymath}
D = \left( \begin{array}{cc}
-1_{9-p} & 0 \\
0 & 1_{p+1}
\end{array}
\right) \ ,\,\, \mand
M_p = \left\{\begin{array}{cc}\frac{\pm i}{(p+1)!}\ga^{i_{1}}\ga^{i_{2}}\ldots \ga^{i_{p+1}}
\eps_{i_{1}\ldots i_{p+1}}\,\,\,\,{\rm for\, p \,even}\\ \frac{\pm 1}{(p+1)!}\ga^{i_{1}}\ga^{i_{2}}\ldots \ga^{i_{p+1}}\ga_{11}
\eps_{i_{1}\ldots i_{p+1}} \,\,\,\,{\rm for\, p \,odd}\end{array}\right.
\non\end{displaymath}

where now the propagators for all world-sheet fields are
\begin{eqnarray}
\lan X^{\mu}(z)X^{\nu}(w)\ran & = & -\frac{\alpha'}{2}\eta^{\mu\nu}\log(z-w) \ , \non \\
\lan \psi^{\mu}(z)\psi^{\nu}(w) \ran & = & -\frac{\alpha'}{2}\eta^{\mu\nu}(z-w)^{-1} \ ,\non \\
\lan\phi(z)\phi(w)\ran & = & -\log(z-w) \ .
\labell{prop2}\end{eqnarray}.}

Applying those vertex operators and Wick theorem, our desired S-matrix can be written down by

\beqa
&&\int dx_1  dx_4 dx_5  (P_{-}\fsC_{(n-1)}M_p)^{\alpha\beta}  (x_{45})^{-3/4}  \xi_{1i}  \bigg(I_1+(2i k_{1a})I_2\bigg) I
 \nonumber\eeqa
so that  $x_4=z=x+iy,x_5=\bar z=x-iy$ and 
\beqa
I&=&|x_{14}x_{15}|^{\frac{\alpha'^2}{2}k_1.p} |x_{45}|^{\frac{\alpha'^2}{4}p.D.p},\label{bb}
\eeqa
with
 \beqa
 I_1= ip^{i}(x_{54})^{-5/4} C^{-1}_{\alpha\beta}\frac{x_{54}}{x_{14}x_{15} }
 \eeqa

It is also important to talk about the following correlator which can be  obtained by generalising the Wick-like rule \cite{Liu:2001qa,Garousi:2008ge}

  \beqa
  I_2= <:S_{\al}(x_4): S_{\be}(x_5):\psi^{a}\psi^{i}(x_1):>&=& 2^{-1}(x_{14}x_{15})^{-1}(x_{45})^{-1/4}(\Gamma^{ia} C^{-1})_{\alpha\beta}
  \nonumber\eeqa

 If we would replace the above correlators inside the amplitude then we could see that our S-matrix does respect the  $SL(2,R)$ invariance. We gauge fix it as  $(x_1,z,\bar z)=(\infty,i,-i)$ so that the final result of  our  S-matrix in this certain picture  is
\beqa
{\cal A}^{\phi^{0},C^{-2}} &=& \bigg[-ip^i\Tr(P_{-}\fsC_{(n-1)}M_p)+ik_{1a}\Tr(P_{-}\fsC_{(n-1)}M_p\Gamma^{ia})\bigg]\xi_{1i} \labell{cc}\ .
\eeqa

As it can be seen from the above S-matrix , it seems to have two different terms in our amplitude while in below we show that some crucial subtleties are needed. Note that after the derivation of the S-matrix, one could start writing all its field theory couplings to be compared with the amplitude while before doing so, we claim that one has to know the correct form of  the S-matrix. Hence let us carry out this amplitude in the other picture  $<C^{(-1)}\phi^{(-1)}>$ in the next section and back to the subtlety associated to \reef{cc} afterwards.

\vskip.2in

It is also worth deriving the S-matrix of one RR and a gauge field in the asymmetric picture of RR.\footnote{ Let us just mention the final answer
\beqa
{\cal A}^{A^{0},C^{-2}} &=& \xi_{1a}
\bigg[-i p^a\Tr(P_{-}\fsC_{(n-1)}M_p)+ik_{1b}\Tr(P_{-}\fsC_{(n-1)}M_p\Gamma^{ab})\bigg]\nonumber
\eeqa

 Note that the first term in above S-matrix has definitely no contribution to the amplitude because if we apply momentum conservation along the world volume of brane $(k_1+p)^a=0$ and on-shell condition for the gauge field gives rise the first part of the S-matrix to be vanished. Indeed the second part of the S-matrix can be produced by $2\pi\alpha'\int C_{p-1}\wedge F $ coupling as the three point function of  a symmetric closed string RR and one gauge field in  (-1) picture was given by
\beqa
V_A^{(-1)}(x)&=&e^{-\phi(x)}\xi_a\psi^a(x)e^{2iq\inn X(x)} \nonumber\\
{\cal A}^{C^{-1}A^{-1}} & \sim & 2^{-1/2}\xi_{1a} \Tr(P_{-}\fsH_{(n)}M_p\gamma^a)
 \nonumber\eeqa}

\vskip.2in

\section{The $C^{-1}- \phi^{-1}$ amplitude}

The three point function from the world-sheet prospective ( two point function from the space time point of view ) of  one closed string RR
and a  real tachyon of type II super string in  their different pictures  has been done.\footnote{\beqa
{\cal A}^{C^{-1}T^{-1}} & \sim & -2i \Tr(P_{-}\fsH_{(n)}M_p)
 \nonumber\eeqa
and also \beqa
{\cal A}^{C^{-2}T^{0}} & \sim & 2^{1/2} (2ik_{1a})\Tr(P_{-}\fsC_{(n-1)}M_p\gamma^a)
 \nonumber\eeqa
now if we apply momentum conservation along the world volume of brane $(k_1+p)^a=0 $, extract the trace and use $p\fsC=\fsH$  (up to normalisation constant) we get the same S-matrix in both pictures, where this S-Matrix can be generated with $2i\pi\alpha'\beta'\mu'_p\int C_p\wedge DT$ coupling in field theory.}

 This three point function from the world-sheet prospective with both RR and scalar field in $(-1)$  picture is given by

\beqa
{\cal A}^{\phi,RR} & \sim & \int dxd^2z
 \lan V_\phi^{(-1)}(x)
V_{RR}^{(-1)}(z,\bar{z})\ran\labell{dd}\eeqa
The vertices are written down so that RR is considered in terms of its field strength in symmetric picture.  They are presented  as follows
\beqa
V_\phi^{(-1)}(x)&=&e^{-\phi(x)}\xi_i\psi^i(x)e^{2iq\inn X(x)} \nonumber\\
V_{RR}^{(-1)}(z,\bar{z})&=&(P_{-}\fsH_{(n)}M_p)^{\al\be}e^{-\phi(z)/2} S_{\al}(z)e^{i\frac{\alpha'}{2}p\cd X(z)}e^{-\phi(\bar{z})/2} S_{\be}(\bar{z}) e^{i\frac{\alpha'}{2}p\cd D \cd X(\bar{z})}\nonumber
\eeqa

 Obviously all the previous definitions of the first section for projector, holomorphic components and the other field contents  have been held here as well.
Once more we substitute the defined vertex operators into \reef{dd} and the amplitude reduces to

\beqa
&&\int dx_1  dx_4 dx_5  (P_{-}\fsH_{(n)}M_p)^{\alpha\beta}(x_{14}x_{15})^{-1/2} (x_{45})^{-1/4}  \xi_{1i}  I
 \nonumber\\&&\times
<:S_{\al}(x_4): S_{\be}(x_5):\psi^{i}(x_1):>
 \nonumber\eeqa


where the result for the following correlation function is needed, that is ,
  \beqa
  <:S_{\al}(x_4): S_{\be}(x_5):\psi^{i}(x_1):>&=& 2^{-1/2}(x_{14}x_{15})^{-1/2}(x_{45})^{-3/4}(\gamma^{i} C^{-1})_{\alpha\beta}
  \nonumber\eeqa

The $SL(2,R)$ invariance of the S-matrix can be readily checked and we did gauge fixing as $(\infty,i,-i)$\footnote{  we set $\alpha'=2$ }. The final result of the S-matrix of one scalar and one RR closed string in this symmetric picture is 
\beqa
{\cal A}^{C^{-1}\phi^{-1}} &=& 2^{-1/2}\Tr
(P_{-}\fsH_{(n)}M_p\gamma^{i})\xi_{1i} \labell{ee}\ .
\eeqa

One should pay particular attention to the the conservation of momentum along the world volume of brane as  $k_1^{a} + p^{a} =0$. Let us first reproduce the field theory of above S-Matrix.  The amplitude might be  normalised by a coefficient of  $ (2^{1/2}\pi\mu_p/8)$ such that $  \mu_p $ is Ramond-Ramond charge of brane. The trace is done as follows\footnote{The trace with  $\gamma^{11}$ can be shown that  all the above results kept held even for the following case
\beqa
  p>3 , H_n=*H_{10-n} , n\geq 5
  \nonumber\eeqa}
\beqa
\Tr\bigg(\fsH_{(n)}M_p
\ga^i\bigg)\delta_{p+2,n}&=&\pm\frac{32}{(p+2)!}\eps^{a_{0}\cdots a_{p}}H_{ia_{0}\cdots a_{p}}
  \delta_{p+2,n}\nonumber\eeqa
  
Eventually this S-matrix  \reef{ee} can be precisely produced with the following field theory coupling
\beqa
    \mu_p (2\pi\alpha')\int_{\Sigma_{p+1}} \bigg(\Tr(\partial_{i}C_{p+1}\phi^i)\bigg)
    \label{jj}\eeqa

where we have used  the so called Taylor expansion of a scalar field, meanwhile in the other picture , the S-matrix  was found to be
\beqa
{\cal A}^{\phi^{0},C^{-2}} &=& \bigg[-i p^i\Tr(P_{-}\fsC_{(n-1)}M_p)+ik_{1a}\Tr(P_{-}\fsC_{(n-1)}M_p\Gamma^{ia})\bigg]\xi_{1i}\nonumber 
\eeqa

Let us compare \reef{cc} with \reef{ee}. We know that $p^i C=H^i$ so up to normalisation constant the first term of \reef{cc} does exactly produce the same  S-matrix of \reef{ee}, therefore  we claim that the second term of \reef{cc} has no contribution to the S-matrix of one scalar and one RR at all.  Hence the prescription for removing or getting rid of the second term of \reef{cc} is as follows.

We first apply the momentum conservation along the world volume of brane to the second term ( $k_1^{a} + p^{a} =0$) and then extract its trace as follows:

\beqa
\Tr\bigg(\fsC_{(n-1)}M_p
\Gamma^{ia}\bigg)\delta_{p+2,n}&=&\pm\frac{32}{(p+1)!}\eps^{a_{0}\cdots a_{p-1}a}C_{ia_{0}\cdots a_{p-1}}
  \delta_{p+2,n}\nonumber\eeqa

 and more importantly in order to  get to the same S-matrix as ${\cal A}^{\phi^{-1},C^{-1}}$, we understand that  the following Bianchi identity must hold for BPS branes

   \beqa
   p^a\eps^{a_{0}\cdots a_{p-1}a} =0
   \eeqa

However, this is not the full story, as we will see in the next sections  for the higher point functions of string amplitudes, one has to generalise all the Bianchi identities \footnote{In particular in order not to miss any contact interactions, one might need to look at the other pictures of the higher point functions of string theory amplitudes as well.}. Let us now turn to some of the four point functions and obtain new Bianchi identities.

\section{The $T^{-1}-\phi^{0}-C^{-1}$ amplitude}

 The four point function from the world-sheet prospective ( three point function from the space time point of view ) of  one closed string RR and two real tachyons of type II super string with  their all different pictures can  been done.\footnote{ Indeed after performing the gauge fixing as  $(x_1,x_2,z,\bar z)=(x,-x,i,-i)$ with $u = -\frac{\alpha'}{2}(k_1+k_2)^2$ the S-matrix is
 \beqa
V_{T}^{(-1)}(y) &=&  e^{-\phi(y)} e^{2ik\cd X(y)}\otimes \sigma_2\nonumber\\&&
{\cal A}^{C^{-1}T^{-1}T^{0}} \sim  2^{3/2} \pi \frac{\Gamma[-2u]}{\Gamma[1/2-u]^2}\Tr(P_{-}\fsH_{(n)}M_p\gamma^a)k_{2a}
 \label{yy1}\eeqa
where one can use momentum conservation along the brane $-k_1^{a} - p^{a} =k_2^{a}$ and apply  the following Bianchi identity $p_a \eps^{a_{0}...a_{p-1}a}=0$, to show that the S-matrix is antisymmetric with respect to interchanging the two tachyons, whereas this S-matrix in asymmetric picture is \beqa
{\cal A}^{C^{-2}T^{0}T^{0}} & \sim & 4 k_{1a}k_{2b}\int_{-\infty}^{\infty} dx (2x)^{-2u-1}
\bigg((1+x^{2})\bigg)^{2 u} \bigg[\Tr
(P_{-}\fsC_{(n-1)}M_p\Gamma^{ba})-2\eta^{ab}\frac{1-x^2}{4xi}\Tr(P_{-}\fsC_{(n-1)}M_p)\bigg],\nonumber\eeqa
where evidently the second term has no contribution to the S-matrix , because the integration must be taken over the whole space-time  and the integrand is odd function so the result for the second term is zero. Now if we apply  momentum conservation to the first term of above S-matrix and more significantly in order to make sense of non-zero S-matrix in this asymmetric picture,  we believe that  the following equation
\beqa
p_a \eps^{a_{0}...a_{p-2}ba}
\nonumber\eeqa
must be non-zero, otherwise the whole S-matrix vanishes.

Notice that all u-channel poles with an infinite higher derivative corrections to     $ (2\pi\alpha')^2\beta'\mu'_p\int C_{p-1}\wedge DT\wedge DT$ coupling can also be derived.
}

 The four point function of one tachyon, a scalar and one RR from the world-sheet point of view has been performed in detail in \cite{Hatefi:2013yxa}. Indeed both the S-matrix  and its field theory part  in the following picture $T^{0}-\phi^{-1}-C^{-1}$ have already been computed. Truly, after carrying out the gauge fixing as  $(x_1,x_2,z,\bar z)=(x,-x,i,-i)$  the S-matrix is given by\footnote { with $u = -\frac{\alpha'}{2}(k_1+k_2)^2$ and ( $k_1^{a} + k_2^{a} + p^{a} =0$).}
\beqa
4k_{1a} \xi_i\int_{-\infty}^{\infty} dx (2x)^{-2u-1/2}
\bigg((1+x^{2})\bigg)^{-1/2 +2 u} \bigg(\Tr
(P_{-}\fsH_{(n)}M_p\Gamma^{ia})\bigg),\nonumber\eeqa

\vskip .2in

     Thus the result of this non super symmetric amplitude is read off as
\beqa
{\cal A}^{T^{0},\phi^{-1},C^{-1}} &=&(\pi\beta'\mu_p')2\sqrt{\pi} \frac{\Ga[-u+1/4]}{\Ga[3/4-u]}
 \Tr
(P_{-}\fsH_{(n)}M_p\Gamma^{ai})k_{1a}\xi_i\Tr(\lam_1\lam_2) \labell{amp383}\ .
\eeqa

where $ (\pi\beta'\mu_p'/2)$ ,$ \beta' $  and   $  \mu_p' $ are defined as normalisation constant,WZ and the RR charge of brane. In effective field theory it was shown that , this S-matrix can be precisely reproduced by the following coupling of type II string theory   \beqa
    2i\beta'\mu'_p (2\pi\alpha')^2\int_{\Sigma_{p+1}} \bigg(\Tr(\partial_{i}C_{p}\wedge DT\phi^i)\bigg),
    \label{jj}\eeqa
and all its infinite corrections were derived in \cite{Hatefi:2013yxa}.\footnote{\beqa
\frac{2\beta'\mu_p'}{p!}(2\pi\alpha')^2\epsilon^{a_{0}...a_{p}} \partial_{i}C_{a_{0}...a_{p-1}}\wedge \Tr\left(\sum_{m=-1}^{\infty}c_m(\alpha')^{m+1}  D_{a_1}\cdots D_{a_{m+1}}D_{a_{p}}T D^{a_1}...D^{a_{m+1}}\phi^i\right) \labell{highaa}\eeqa.}
Let us carry it out in the other picture so if we use the above vertex operators and perform  all the correlators with same techniques that have been explained  in the previous section, then one explores the final form of the S-matrix in this picture as
 \beqa
{\cal A}^{C^{-1}\phi^{0}T^{-1}} & \sim & 4 \xi_{1i}\int_{-\infty}^{\infty} dx (2x)^{-2u-1/2}
\bigg((1+x^{2})\bigg)^{2u-1/2} \nonumber\\&&\times\bigg[k_{1a}\Tr
(P_{-}\fsH_{(n)}M_p\Gamma^{ia})-p^i\Tr(P_{-}\fsH_{(n)}M_p)\bigg],\label{ww12}\eeqa

Now in order to get to the same S-matrix element for this four point world sheet amplitude as appeared in \reef{amp383}, 
one has to apply the momentum conservation $(k_1+k_2+p)^a=0$ and keep in mind the following Bianchi identity as well.

\beqa
p^i \eps^{a_{0}...a_{p}}H_{a_{0}...a_{p}}+p^a \eps^{a_{0}...a_{p-1}a}H^{i}_{a_{0}...a_{p-1}}&=&0
\label{rr44}\eeqa

\vskip.1in

Finally let us calculate this S-matrix  in asymmetric picture of RR and make some essential comments about this four point function.

\vskip.2in

Notice that there is a non zero  coupling between two gauge fields and one RR in the world volume of BPS branes of type II string theory  .\footnote{ In \cite{Hatefi:2011jq} it is shown that

\beqa
{\cal A}^{C^{-1}A^{-1}A^{0}} & \sim & 2^{-3/2} \xi_{1a}\xi_{2b}\int_{-\infty}^{\infty} dx (2x)^{-2u}
\bigg((1+x^{2})\bigg)^{2u-1} \bigg\{4 k_{2c}\Tr(P_{-}\fsH_{(n)}M_p\Gamma^{bca})+(\frac{1-x^2}{x})\bigg[k_{1b}\Tr
(P_{-}\fsH_{(n)}M_p\gamma^{a})\nonumber\\&&+k_{2c}\bigg(-\eta^{ac}\Tr(P_{-}\fsH_{(n)}M_p\gamma^b)+\eta^{ab}\Tr(P_{-}\fsH_{(n)}M_p\gamma^c)\bigg)\bigg]\bigg\}\nonumber\eeqa
where obviously the last three terms of the S-matrix have no no contribution to amplitude, because the integration must be taken over the whole space and the integrand is odd.}
\section{The $T^{0}-\phi^{0}-C^{-2}$ amplitude}

The four point function of an asymmetric  RR, a scalar and an open string tachyon can be investigated by the following correlation function 

\beqa
{\cal A}^ {T^{0}\phi^{0}C^{-2}} & \sim & \int dx_1 dx_2 d^2z
 \lan V_\phi^{(0)}(x_1)V_T^{(0)}(x_2)
V_{RR}^{(-2)}(z,\bar{z})\ran\labell{cor102}\eeqa
where the tachyon, scalar field and RR vertex operators  are  given as \footnote {To see the Chan-Paton factors see \cite{Hatefi:2013yxa} }

\beqa
V_{T}^{(0)}(y) &=&\alpha'ik_1\cd\psi(y) e^{\alpha'ik_1\cd X(y)}\lam\otimes\sigma_1 \nonumber\\
V_{\phi}^{(0)}(x) &=& \xi_{1i}(\partial^i X(x)+i\alpha'k.\psi\psi^i(x))e^{\alpha'ik.X(x)}\lam\otimes I \nonumber\\
V_{RR}^{(-2)}(z,\bar{z})&=&(P_{-}\fsC_{(n-1)}M_p)^{\al\be}e^{-3\phi(z)/2} S_{\al}(z)e^{i\frac{\alpha'}{2}p\cd X(z)}e^{-\phi(\bar{z})/2} S_{\be}(\bar{z}) e^{i\frac{\alpha'}{2}p\cd D \cd X(\bar{z})}\lam\otimes\sigma_3\sigma_1\nonumber
\eeqa

Applying Wick theorem ,the  amplitude  can be explored as follows
\beqa
&&\int dx_1 dx_2 dx_4 dx_5  (P_{-}\fsC_{(n-1)}M_p)^{\alpha\beta}  (x_{45})^{-3/4} (4ik_{2a}) \xi_{1i}  \bigg(I_3+(2i k_{1c})I_4\bigg) I_5
 \nonumber\eeqa
with 
\beqa
I_5&=&|x_{12}|^{\alpha'^2 k_1.k_2}|x_{14}x_{15}|^{\frac{\alpha'^2}{2}k_1.p} |x_{24}x_{25}|^{\frac{\alpha'^2}{2}k_2.p}|x_{45}|^{\frac{\alpha'^2}{4}p.D.p},\nonumber\\
 I_3&=&ip^{i} 2^{-1/2} (x_{24}x_{25})^{-1/2} (x_{45})^{-3/4} (\gamma^a C^{-1})_{\alpha\beta}\frac{x_{54}}{x_{14}x_{15} }
 \eeqa

  Now using Wick-like rule one gets to the following correlator
  \beqa
  I_4&=& <:S_{\al}(x_4): S_{\be}(x_5):\psi^{c}\psi^{i}(x_1):\psi^{a}(x_2):>= 2^{-3/2}(x_{24}x_{25})^{-1/2}(x_{14}x_{15})^{-1}(x_{45})^{1/4} \nonumber\\&& \bigg[(\Gamma^{aic} C^{-1})_{\alpha\beta} +2 \frac{Re[x_{14}x_{25}]}{x_{12}x_{45}}\eta^{ac}(\gamma^{i} C^{-1})_{\alpha\beta}\bigg]
  \nonumber\eeqa

 By applying the above correlators  into this four point amplitude we can easily observe that the integrand or S-matrix is $SL(2,R)$ invariant.
 We do the proper gauge fixing  as  $(x_1,x_2,z,\bar z)=(x,-x,i,-i)$, taking $u=- \frac{\alpha'}{2} (k_1+k_2)^2$ we  obtain the S-matrix as
 \beqa
 {\cal A}^{T^{0}\phi^{0}C^{-2}}={\cal A}_1+{\cal A}_2 \nonumber\eeqa

 such that
\beqa
{\cal A}_{1}^{T^{0}\phi^{0}C^{-2}} &=& 2^{3/2}\xi_{1i}  k_{2a} p^i \Tr\bigg(\fsC_{(n-1)}M_p
\gamma^{a}\bigg) \int_{-\infty}^{\infty} (2x)^{-2t-1/2} (x^2+1)^{2t-1/2}\nonumber\\&&
= 2^{3/2}\xi_{1i}  4k_{2a} p^i \Tr\bigg(\fsC_{(n-1)}M_p
\gamma^{a}\bigg)\sqrt{\pi} \frac{\Ga[-u+1/4]}{\Ga[3/4-u]}
\eeqa

Now if we use the momentum conservation $(k_1+k_2+p)^a=0$ and the Bianchi identity $p^a\eps^{a_{0}\cdots a_{p-1}a} =0$ ,then we come to the point that up to a coefficient of $2^{3/2}$ this part of the S-matrix exactly produces \reef{amp383}. The second part of the S-matrix is found out to be
\beqa
{\cal A}_{2}^{T^{0}\phi^{0}C^{-2}} &=& 2^{3/2}\xi_{1i} k_{1c} k_{2a}  \int_{-\infty}^{\infty} (2x)^{-2t-1/2} (x^2+1)^{2t-1/2} \bigg[2\eta^{ac}\Tr\bigg(\fsC_{(n-1)}M_p
\gamma^{i}\bigg)\frac{1-x^2}{4xi}\nonumber\\&&+\Tr\bigg(\fsC_{(n-1)}M_p
\Gamma^{aic}\bigg)\bigg]\label{ff}
\eeqa

   where the first term in \reef{ff} is indeed zero because, the integration is taken over the whole space while the integrand is odd so naturally the answer for the  first term of \reef{ff} is zero\footnote{
    \beqa
 2^{3/2}\xi_{1i} (-u-1/4) \Tr\bigg(\fsC_{(n-1)}M_p
\gamma^{i}\bigg) \int_{-\infty}^{\infty} \bigg(\frac{(1+x^2)^2}{(4x^2)}\bigg)^{1/4+u} \frac{1-x^2}{(x^2+1)(4xi)} =0
\eeqa.}

However, as it is clear the result for the second term of \reef{ff} is non-zero, that is,

\beqa
{\cal A}_{2}^{T^{0}\phi^{0}C^{-2}} &=& 2^{3/2}\xi_{1i} 4k_{1c} k_{2a}  \Tr(\fsC_{(n-1)}M_p\Gamma^{aic})\sqrt{\pi} \frac{\Ga[-u+1/4]}{\Ga[3/4-u]}\label{ffl}
\eeqa

Therefore we might think of this term as the extra contact interaction to the S-matrix , however, after applying momentum conservation along the world volume of brane and using Bianchi identities \footnote{$p^a\eps^{a_{0}\cdots a_{p-2}ca} =p^c\eps^{a_{0}\cdots a_{p-2}ca}=0$}, we reveal the fact that this term has zero contribution to the  S-matrix of an asymmetric RR, a scalar and a tachyon. \footnote{ It is important to point out that this term by itself without applying any Bianchi identity equation could be meant to be  non-zero and might have been confused that it plays the role of the whole infinite contact interactions/ surface terms or total derivatives  where clearly it does not play any contribution to the whole S-matrix
.}

\section{The $\phi^{0}-A^{0}-C^{-2}$ amplitude}

The four point function of an  asymmetric RR, a scalar and a gauge field  can be carried out by the following correlation function,
\beqa
{\cal A}^ {\phi^{0} A^{0}C^{-2}} & \sim & \int dx_1 dx_2 d^2z
 \lan V_\phi^{(0)}(x_1)V_A^{(0)}(x_2)
V_{RR}^{(-2)}(z,\bar{z})\ran\labell{cor10ti}\eeqa
where the scalar field and RR vertex operators  are  given in the previous sections, and for the gauge field we have
\beqa
V_{A}^{(0)}(x) &=& \xi_{2a}(\partial^a X(x)+i\alpha'k.\psi\psi^a(x))e^{\alpha'ik.X(x)}
\nonumber
\eeqa

Having set the Wick theorem , the  amplitude  may have been written down as
\beqa
&&\int dx_1 dx_2 dx_4 dx_5  (P_{-}\fsC_{(n-1)}M_p)^{\alpha\beta}  (x_{45})^{-3/4} \xi_{1i}\xi_{2a}  \bigg(J_1+J_2+J_3+J_4\bigg) I_5
 \nonumber\eeqa
and also 
\beqa
I_5&=&|x_{12}|^{\alpha'^2 k_1.k_2}|x_{14}x_{15}|^{\frac{\alpha'^2}{2}k_1.p} |x_{24}x_{25}|^{\frac{\alpha'^2}{2}k_2.p}|x_{45}|^{\frac{\alpha'^2}{4}p.D.p},\nonumber\\
\eeqa
 where by applying the generalization of Wick-like rule one can obtain all the correlators as follows

\beqa
 J_1&=& ip^{i}\frac{x_{54}}{x_{14}x_{15}} (l_a)   (x_{45})^{-5/4} ( C^{-1})_{\alpha\beta}\nonumber\\&&
l_a = -ik_{1a}\bigg[\frac{x_{14}}{x_{12}x_{24}}+\frac{x_{15}}{x_{12}x_{25}}\bigg]\nonumber\\&&
J_2 =-p^{i}k_{2c}\frac{x_{54}}{x_{14}x_{15}}  (x_{24}x_{25})^{-1}  (x_{45})^{-1/4} ( \Gamma^{ac}C^{-1})_{\alpha\beta}\nonumber\\&&
 J_3= (l_a) ik_{1b} (x_{14}x_{15})^{-1}  (x_{45})^{-1/4} ( \Gamma^{ib}C^{-1})_{\alpha\beta}\nonumber\\&&
 J_4=-k_{1b}k_{2c} (x_{14}x_{15}x_{24}x_{25})^{-1}  (x_{45})^{3/4} \nonumber\\&&\times \bigg[ ( \Gamma^{acib}C^{-1})_{\alpha\beta}+(2\eta^{bc}(\Gamma^{ai}C^{-1})_{\alpha\beta}-2\eta^{ab}(\Gamma^{ci}C^{-1})_{\alpha\beta})\frac{Re[x_{14}x_{25}]}{x_{12}x_{45}}\bigg] \label{mion}\eeqa

If we now apply \reef{mion}  into this four point amplitude we can easily determine that the  S-matrix is $SL(2,R)$ invariant. We do the proper gauge fixing  as  $(x_1,x_2,z,\bar z)=(x,-x,i,-i)$, taking $t=- \frac{\alpha'}{2} (k_1+k_2)^2$ to actually get to the entire S-matrix as

 \beqa
 {\cal A}^{\phi^{0}A^{0}C^{-2}}&=&-\xi_{1i}\xi_{2a}\int_{-\infty}^{\infty}dx (1+x^2)^{2t-1} (2x)^{-2t}
 \bigg[\frac{1-x^2}{x}\bigg(-ip^i k_{1a}\Tr(P_{-}\fsC_{(n-1)}M_p)\nonumber\\&&+k_{1b}k_{1a}\Tr(P_{-}\fsC_{(n-1)}M_p\Gamma^{ib})
 +\eta^{bc}\Tr(P_{-}\fsC_{(n-1)}M_p\Gamma^{ai})
 -\eta^{ab}\Tr(P_{-}\fsC_{(n-1)}M_p\Gamma^{ci})\bigg)\nonumber\\&&
 +2ik_{2c}p^i\Tr(P_{-}\fsC_{(n-1)}M_p\Gamma^{ac})
 -2ik_{1b}k_{2c}\Tr(P_{-}\fsC_{(n-1)}M_p\Gamma^{acib})\bigg]\label{lop}\eeqa

\vskip.2in

where the first, second, third and fourth term do not have any contribution to the S-matrix because the integration is taken on the whole space and the integrand is odd. 
 If we use the momentum conservation $(k_1+k_2+p)^a=0$ and the Bianchi identity $p_b\eps^{a_{0}\cdots a_{p-3}bac} =0$ , then we come to the point that  the sixth term has also no contribution to the amplitude so only the fifth term has non-zero contribution to the S-matrix of an asymmetric RR, a scalar and a gauge field.

Hence, the final result is

 \beqa
 {\cal A}^{\phi^{0}A^{0}C^{-2}}&=&-\xi_{1i}\xi_{2a} 2ik_{2c}p^i\Tr(P_{-}\fsC_{(n-1)}M_p\Gamma^{ac}) \pi^{1/2}\frac{\Ga[-t+1/2]}{\Ga[1-t]}\label{nnb2}\eeqa

where the expansion of the amplitude is non-zero for $p=n$ case and it does not include any poles as it is clear from the \reef{nnb2}, because the low energy expansion is $t\rightarrow 0$ limit and all the infinite contact interactions of this S-matrix have already been derived in \cite{Hatefi:2012ve}. Given the closed form of the above correlation functions one can find out to 
${\cal A}^{\phi^{-1}A^{0}C^{-1}}$ as as well

 \beqa
 {\cal A}^{\phi^{-1}A^{0}C^{-1}}&=&2^{-3/2}\xi_{1i}\xi_{2a}\int_{-\infty}^{\infty}dx (1+x^2)^{2t-1} (2x)^{-2t}
 \bigg[\frac{1-x^2}{x}\bigg(k_{1a}\Tr(P_{-}\fsH_{(n)}M_p\gamma^i)\bigg)\nonumber\\&&
 -2ik_{2b}\Tr(P_{-}\fsH_{(n)}M_p\Gamma^{abi})\bigg]\nonumber\eeqa
where again the first term has no contribution and the second term 
 (up to a coefficient of $2^{3/2}$) precisely produces \reef{nnb2}. Finally one explores this amplitude in its last picture as follows

 \beqa
 {\cal A}^{\phi^{0}A^{-1}C^{-1}}&=&2^{1/2}\xi_{1i}\xi_{2a}\int_{-\infty}^{\infty}dx (1+x^2)^{2t-1} (2x)^{-2t}
 \bigg[k_{1b}\Tr(P_{-}\fsH_{(n)}M_p\Gamma^{bai})\nonumber\\&&-p^i\Tr(P_{-}\fsH_{(n)}M_p\gamma^{a})\bigg]\label{mty}\eeqa

 where in \reef{mty}, one has to apply momentum conservation to its first term and use the following Bianchi identity 
 \beqa
 p_b\eps^{a_{0}\cdots a_{p-2}ba} H^i_{a_{0}\cdots a_{p-2}}+p^i\eps^{a_{0}\cdots a_{p-1}a} H_{a_{0}\cdots a_{p-1}} &=&0
 \label{majid}\eeqa
to actually get to the entire S-matrix as appeared in \reef{nnb2}.

 Therefore in this particular picture , again we are just  left with one term for the final answer of the a RR, a gauge and a scalar field and this term is necessary because this S-matrix has to be produced by a non-zero coupling  $(2\pi\alpha')^2\mu_p\int _{H^{+}}\partial_{i}C_{a_{0}..a_{p-2}} F_{a_{p-1}a_{p}}  \phi^i$ of effective field theory  where the scalar field 
  comes from Taylor expansion.

Note that by comparing this S-matrix with field theory, we come to understand that there should not be any other term  in effective field theory coming from the pull-back of brane.

\vskip.2in

 Now in order to obtain the other non trivial Bianchi identities, we are going to consider one of the most simplest five point functions and deal with more subtleties about perturbative string theory.


 \section{ The five point world-sheet S-matrix of brane-anti brane system }

It is known that the world volume of brane-anti brane system must have two real tachyon fields.\footnote{
In \cite{Michel:2014lva} it is discussed in detail that  brane anti brane system should be investigated by means of the effective field theory techniques which seems to be the proper way of realising classical or even for loop divergences for which anti brane dynamics in the presence of some of background fields  may play a major role.} The complete form of the amplitude of a gauge ,two real tachyons and a closed string RR of  brane anti brane for various $p,n$ cases  $\lan V_{A}^{(-1)}{(x_{1})}
V_{T}^{(0)}{(x_{2})}V_T^{(0)}{(x_{3})}
V_{RR}^{(-\frac{1}{2},-\frac{1}{2})}(z,\bar{z})\ran$ has been derived in \cite{Garousi:2007fk}.
 \vskip.1in

For the completeness we have done this amplitude in the following picture as well
 
\beqa
\lan V_{A}^{(0)}{(x_{1})}
V_{T}^{(-1)}{(x_{2})}V_T^{(0)}{(x_{3})}
V_{RR}^{(-\frac{1}{2},-\frac{1}{2})}(z,\bar{z})\ran \nonumber\eeqa 

and the final result for the  amplitude is exactly the same as appeared in \cite{Garousi:2007fk} so that  it satisfies the Ward identity associated to the gauge field. It is also worth explaining that the S-matrix elements of one closed string RR field ,one scalar field and two tachyons
on the world volume brane anti brane system in the following picture has also been computed in detail in
 \cite{Hatefi:2012cp}\footnote{ The following vertices with their correct Chan-Paton factors of D-brane anti D-brane  are held 
 \beqa
V_{\phi}^{(-1)}(x) &=& \xi_{i}\psi^i(x)e^{2iq.X(x)}e^{-\phi(x)}\otimes \sigma_3
\nonumber\\
V_{T}^{(0)}(y) &=& 2ik.\psi(y)  e^{2ik\cd X(y)},
\labell{vertex1}\otimes \sigma_1\\
V_{RR}^{(-\frac{1}{2},-\frac{1}{2})}(z,\bar{z})&=&(P_{-}\fsH_{(n)}M_p)^{\al\be}e^{-\phi(z)/2}
S_{\al}(z)e^{ip\cd X(z)}e^{-\phi(\bar{z})/2} S_{\be}(\bar{z})
e^{ip\cd D \cd X(\bar{z})}\otimes \sigma_3,\nonumber\eeqa
so that  $k^2=1/4$ is the condition for tachyons in type II string theory and the following definitions for Mandelstam variables are used
\beqa
s&=&-\frac{\alpha'}{2}(k_1+k_3)^2, t=-\frac{\alpha'}{2}(k_1+k_2)^2, u=-\frac{\alpha'}{2}(k_2+k_3)^2,\labell{man}\eeqa
}

 \beqa
\lan V_{\phi}^{(-1)}{(x_{1})}
V_{T}^{(0)}{(x_{2})}V_T^{(0)}{(x_{3})}
V_{RR}^{(-\frac{1}{2},-\frac{1}{2})}(z,\bar{z})\ran\label{tt}\eeqa

The final form of this S-matrix in this particular picture  was given as
\beqa {\cal A}^{C^{(-1)}\phi^{(-1)} T^{(0)} T^{(0)}}&=&{\cal A}_{1}+{\cal A}_{2}\labell{181u}\eeqa
where
\beqa
{\cal A}_{1}&\!\!\!\sim\!\!\!&-8\xi_{1i}k_{2a}k_{3b} 2^{-3/2}
\Tr(P_{-}\fsH_{(n)}M_p\Gamma^{bai}
)L_1,
\nonumber\\
{\cal A}_{2}&\sim&8\xi_{1i} 2^{-3/2}\bigg\{\Tr(P_{-}\fsH_{(n)}M_p \gamma^{i})\bigg\}L_2
\labell{4859}\eeqa
where
 $L_1,L_2$ are written down in below
   just in terms of  Gamma functions ( no hypergeometric function is needed)
\beqa
L_1&=&(2)^{-2(t+s+u)-1}\pi{\frac{\Gamma(-u)
\Gamma(-s+\frac{1}{4})\Gamma(-t+\frac{1}{4})\Gamma(-t-s-u)}
{\Gamma(-u-t+\frac{1}{4})\Gamma(-t-s+\frac{1}{2})\Gamma(-s-u+\frac{1}{4})}},\label{kk43}\\
L_2&=&(2)^{-2(t+s+u+1)}\pi{\frac{\Gamma(-u+\frac{1}{2})
\Gamma(-s+\frac{3}{4})\Gamma(-t+\frac{3}{4})\Gamma(-t-s-u-\frac{1}{2})}
{\Gamma(-u-t+\frac{1}{4})\Gamma(-t-s+\frac{1}{2})\Gamma(-s-u+\frac{1}{4})}}
\nonumber\eeqa

where in  \cite{Hatefi:2012cp} all the infinite $u-$ channel gauge poles of $L_1$ and $t+s'+u'$ channel scalar poles of this S-matrix  have been precisely produced,  in addition to that all the infinite higher derivative corrections to two scalars-two tachyons of world volume of brane anti brane  without any ambiguity have also been discovered.

 Now let us deal with this S-matrix in the other pictures of both closed-open strings to see what happens to the complete form of the S-matrix and also explore all its contact interactions.



\section{  $ C^{(-2)}  \phi^{(0)}  T^{(0)} T^{(0)} $ }

The S-matrix element of an asymmetric  closed string RR field, one scalar field and two tachyons
on the world volume of brane anti brane system can be found  as follows. \footnote{ The following vertices with their correct Chan-Paton factor for D-brane-anti D-brane are taken into account ,
 \beqa
V_{\phi}^{(0)}(x) &=& \xi_{1i}(\partial^i X(x)+i\alpha'k.\psi\psi^i(x))e^{\alpha'ik.X(x)}\lam\otimes I
\nonumber\\
V_{T}^{(0)}(y) &=& 2ik.\psi(y)  e^{2ik\cd X(y)},
\nonumber\otimes \sigma_1\\
V_{RR}^{(-\frac{3}{2},-\frac{1}{2})}(z,\bar{z})&=&(P_{-}\fsC_{(n-1)}M_p)^{\al\be}e^{-3\phi(z)/2}
S_{\al}(z)e^{ip\cd X(z)}e^{-\phi(\bar{z})/2} S_{\be}(\bar{z})
e^{ip\cd D \cd X(\bar{z})}\otimes I,\label{ii}\eeqa
with  the same definitions for Mandelstam variables kept held.
} Replacing the vertex operators  and performing all the correlators by using the Wick theorem, one can investigate to obtain the whole amplitude as follows

\beqa {\cal A}^{C^{(-2)}  \phi^{(0)}  T^{(0)} T^{(0)}}&\sim& \int
 dx_{1}dx_{2}dx_{3}dx_{4} dx_{5}\,
(P_{-}\fsC_{(n-1)}M_p)^{\al\be}\xi_{1i}x_{45}^{-3/4}( -8 k_{2b}k_{3c}) I \nonumber\\&&\times
\bigg((ip^i \frac{x_{54}}{x_{15}x_{14}})I_1^{cb}+2i k_{1a}I_2^{cbia}\bigg)\label{ll4}\eeqa where
$x_{ij}=x_i-x_j$,
\beqa
I&=&|x_{12}|^{4k_1.k_2}|x_{13}|^{4k_1.k_3}|x_{14}x_{15}|^{2k_1.p}|x_{23}|^{4k_2.k_3}|x_{24}x_{25}|^{2k_2.p}
|x_{34}x_{35}|^{2k_3.p}|x_{45}|^{p.D.p}\label{ll}
\eeqa

Note that we have already generalised  the Wick theorem to get to the femionic correlations in the presence of currents so that one can find the following correlators as
\beqa
I_1^{cb}=<:S_{\al}(x_4):S_{\be}(x_5):\psi^b(x_2)::\psi^c(x_3):>
=2^{-1}x_{45}^{-1/4} (x_{24}x_{25}x_{34}x_{35})^{-1/2}\nonumber\\
\times\bigg((\Gamma^{cb}C^{-1})_{\alpha\beta}-2\eta^{bc}\frac{Re[x_{24}x_{35}]}{x_{23}x_{45}}(C^{-1})_{\alpha\beta}\bigg)
\nonumber
\label{61rne}\eeqa

  also

\beqa
I_2^{cbia}&=&<:S_{\al}(x_4):S_{\be}(x_5):\psi^a \psi^i(x_1):\psi^b(x_2)::\psi^c(x_3):>
=2^{-2}x_{45}^{3/4} (x_{24}x_{25}x_{34}x_{35})^{-1/2} (x_{14}x_{15})^{-1}\nonumber\\&&\bigg\{(\Gamma^{cbia}C^{-1})_{\alpha\beta}
+2\eta^{ab}\frac{Re[x_{14}x_{25}]}{x_{12}x_{45}}(\Gamma^{ci}C^{-1})_{\alpha\beta}
-2\eta^{ac}\frac{Re[x_{14}x_{35}]}{x_{13}x_{45}}(\Gamma^{bi}C^{-1})_{\alpha\beta}\nonumber\\&&
-2\eta^{bc}\frac{Re[x_{24}x_{35}]}{x_{23}x_{45}}(\Gamma^{ia}C^{-1})_{\alpha\beta}
\bigg\}
\nonumber
\label{6mrne}\eeqa

If we substitute all the above correlations inside the amplitude, then one can  show that the property of
SL(2,R) invariance is being investigated. Here we are working with five point function and to our knowledge the best gauge fixing for this amplitude is to fix the locations of all three open strings as follows

\beqar
 x_{1}=0 ,\qquad x_{2}=1,\qquad x_{3}\rightarrow \infty,
 \eeqar
 If we do so, then we get the entire form of the S-matrix in terms of some integrations  on the upper half plane so that the following integrations for various cases need to be performed \beqa
 \int d^2 \!z |1-z|^{a} |z|^{b} (z - \bar{z})^{c}
(z + \bar{z})^{d},
 \eeqa
with $d=0,1,2$ and $a,b,c$ must be just given in terms of the Mandelstam variables. Notice to the point that the result of the above integrations for $d=0,1$ is got from \cite{Fotopoulos:2001pt,Garousi:2007fk} while for  for $d=2$ the result is given in \cite{Hatefi:2012wj}.


If we do gauge fixing, make use of various pure algebraic simplifications  and most particularly  make use of the results of the  integrals that pointed out in \cite{Fotopoulos:2001pt,Garousi:2007fk,Hatefi:2012wj} then we can write down the final result of the amplitude  \reef{ll4} in this asymmetric picture as follows

\beqa {\cal A}^{C^{-2}\phi^{0} T^{0} T^{0}}&=&{\cal A}_{1}+{\cal A}_{2}+{\cal A}_{3}+{\cal A}_{4}\labell{1m81u}\eeqa
where
\beqa
{\cal A}_{1}&\!\!\!\sim\!\!\!&ip^i\xi_{1i}(4 k_{2b}k_{3c})
\Tr(P_{-}\fsC_{(n-1)}M_p\Gamma^{cb})L_1,
\nonumber\\
{\cal A}_{2}&\sim&4i p^i \xi_{1i} \bigg\{\Tr(P_{-}\fsC_{(n-1)}M_p)\bigg\}  L_2
\nonumber\\
{\cal A}_{3}&\sim&-4i \xi_{1i} k_{1a}k_{2b}k_{3c} \bigg\{\Tr(P_{-}\fsC_{(n-1)}M_p \Gamma^{cbia})\bigg\} L_1
\nonumber\\
{\cal A}_{4}&\sim&4i  \xi_{1i} \bigg\{\Tr(P_{-}\fsC_{(n-1)}M_p\Gamma^{bi})\bigg\}  L_2 (k_{1b}+k_{2b}+k_{3b})
\labell{mm2}\eeqa
where the functions
 $L_1,L_2$ are given in \reef{kk43}.

Note that here  in \reef{mm2} we have  dealt with the five point world sheet scattering of brane anti brane in asymmetric picture and in this particular picture we found the terms that  carry momentum of RR in transverse directions.  These terms   are no longer vanished and indeed these terms  potentially have something to do with some of the issues on the perturbative string theory on upper half plane . We believe that these terms are  related to taking integration on different moduli space that have been pointed out in series of papers appeared in \cite{Witten:2012bh}.

\vskip.1in

Now if we compare the S-matrix in this asymmetric picture \reef{mm2} with \reef{4859} , then we might think of the fact that the last two terms of \reef{mm2} are extra singularities. Indeed if we do not apply the Bianchi identity and momentum conservation along the brane to \reef{mm2} as a matter of fact these two terms would be extra terms by themselves. However, if we compare  \reef{mm2} with \reef{4859}, simultaneously extract the trace in the last term of \reef{mm2} and make use of the momentum conservation along the world volume of brane as follows

\beqa
(k_1+k_2+k_3)^a&=&-p^a \label{zz1}\eeqa

we come to the conclusion  that the last term in \reef{mm2} is apparent singularity and this should be removed, in the other words  upon applying the  following Bianchi identity  

\beqa
p^b\eps^{a_{0}\cdots a_{p-1}b}C^{i}_{a_{0}\cdots a_{p-1}}=0
\label{xx1}\eeqa
the last term of \reef{mm2} vanishes.

Note that in below we show that the above Bianchi identity can not be applied to the correlators 

\beqa
\lan V_{T}^{(0)}{(x_{1})}V_{T}^{(0)}{(x_{2})}V_T^{(0)}{(x_{3})} V_{RR}^{(-\frac{3}{2},-\frac{1}{2})}(z,\bar{z})\ran\nonumber \eeqa of non-BPS branes. Indeed after gauge fixing as $(x_1,x_2,z,\bar{z})=(0,1,\infty,z,\bar{z})$ we gain the following non-BPS amplitude

\beqa {\cal A}^{C^{-2}T^{0} T^{0} T^{0}}&=& -16i  (2^{-3/2} k_{1a} k_{2b} k_{3c})(P_{-}\fsC_{(n)}M_p)^{\alpha\beta}
\int d^2z |1-z|^{2t+2u} |z|^{2t+2s} (z - \bar{z})^{-2(t+s+u+1)}
\nonumber\\&&\times \bigg[(\Gamma^{cba} C^{-1})_{\alpha\beta}+\frac{(z + \bar{z})}{2(z - \bar{z})}\bigg((2\eta^{ac}(\gamma^b C^{-1})_{\alpha\beta})-(2\eta^{ab}(\gamma^c C^{-1})_{\alpha\beta})-(2\eta^{bc}(\gamma^a C^{-1})_{\alpha\beta})\bigg)\nonumber\\&&+(2\eta^{bc}(\gamma^a C^{-1})_{\alpha\beta})\frac{1}{(z - \bar{z})}+(2\eta^{ab}(\gamma^c C^{-1})_{\alpha\beta})\frac{|z|^{2}}{(z - \bar{z})}\bigg]
\nonumber\eeqa

We make use of various pure algebraic simplifications  to write down the final result for the above amplitude  in this asymmetric picture as follows

\beqa {\cal A}^{C^{-2}T^{0} T^{0} T^{0}}&=&{\cal A}_{1}+{\cal A}_{2}\nonumber\eeqa
where
\beqa
{\cal A}_{1}&\!\!\!\sim\!\!\!&-16i  (2^{-3/2} k_{1a} k_{2b} k_{3c})\Tr(P_{-}\fsC_{(n-1)}M_p\Gamma^{cba})N_1
\nonumber\\
{\cal A}_{2}&\sim&  -16i (2^{-3/2})\Tr(P_{-}\fsC_{(n-1)}M_p\gamma^{a})N_2 (k_{1a}+k_{2a}+k_{3a})
\nonumber\eeqa
where the functions
 $N_1,N_2$ are given as

 \beqa
N_1&=&(2)^{-2(t+s+u+1)}\pi{\frac{\Gamma(-u)
\Gamma(-s)\Gamma(-t)\Gamma(-t-s-u-\frac{1}{2})}
{\Gamma(-u-t)\Gamma(-t-s)\Gamma(-s-u)}},\label{kk}\\
N_2&=&(2)^{-2(t+s+u)-3}\pi{\frac{\Gamma(-u+\frac{1}{2})
\Gamma(-s+\frac{1}{2})\Gamma(-t+\frac{1}{2})\Gamma(-t-s-u-1)}
{\Gamma(-u-t)\Gamma(-t-s)\Gamma(-s-u)}}
\nonumber\eeqa
 Note that if we use momentum conservation $p^a=-(k_1+k_2+k_3)^a$ for both first and second part of the above non-BPS S-matrix we get
 \beqa
 p^a\eps^{a_{0}\cdots a_{p-3}cba}C_{a_{0}\cdots a_{p-3}}, p^c\eps^{a_{0}\cdots a_{p-3}cba}C_{a_{0}\cdots a_{p-3}},p^a\eps^{a_{0}\cdots a_{p-1}a}C_{a_{0}\cdots a_{p-1}}\nonumber\eeqa
  are not zero.  In particular in order to make sense of   non supersymmetric  amplitudes in the world volume of   non-BPS branes, the following equations 
\beqa
 p^a\eps^{a_{0}\cdots a_{p-3}cba},\quad\quad p^a\eps^{a_{0}\cdots a_{p-1}a}
 \label{vov}\eeqa must be non zero.
 
 \vskip.3in

Hence  for non  super symmetric amplitudes first  we must extract the traces and keep in mind the above points , that is , the equations that we found for some of the BPS branes, can not be applied to non-BPS amplitudes in the presence of  scalar field and closed string RR. Indeed we expect to see that behaviour because the equations that hold for BPS cases not necessary can be held for non supersymmetric cases whereas for BPS branes the equations seem to be more manifest while this may have been changed  after symmetry breaking.
The lesson is that for scattering of the mixed scalars- tachyons in the presence of RR ( in its asymmetric picture), one needs to break several identities that necessary  hold for BPS branes.

\vskip.3in

 What about the third term of \reef{mm2}? One might add it with the first term of \reef{mm2} and come to the point that

  \beqa
  k_{2b}k_{3c}\xi_{1i}(p^i\eps^{a_{0}\cdots a_{p-2}cb}C_{a_{0}\cdots a_{p-2}}+p^a\eps^{a_{0}\cdots a_{p-3}cba}C^{i}_{a_{0}\cdots a_{p-3}})\nonumber\eeqa
   should be vanished, however from  \reef{4859} we know that the first term of \reef{mm2}  holds and plays the crucial role in effective field theory. Thus we need to explore new Bianchi identity for the third term of \reef{mm2}. In fact if we  actually apply momentum conservation to the third term of \reef{mm2} and because of the antisymmetric property of $\eps$ we conclude that the equation $ p^a\eps^{a_{0}\cdots a_{p-3}cba}C^{i}_{a_{0}\cdots a_{p-3}}$ must be vanished for brane anti brane  amplitudes.
 
 \vskip.2in
   
     Therefore, the lesson we have got is as follows. In the presence of an asymmetric RR , a scalar and some tachyons, one needs to find out new Bianchi identities to get to the same S-matrix elements as obtained by a symmetric RR , a scalar and some tachyons. Because we have no gauge field to check the gauge invariance of the amplitude  and more accurately there is no Ward identity for the scalar field .

\vskip.3in

Now let us come over to  the same S-matrix of a symmetric RR, two tachyons and a scalar field in zero picture $< C^{(-1)}  \phi^{(0)}  T^{(-1)} T^{(0)} >$ .

 \section{  $ C^{(-1)}  \phi^{(0)}  T^{(-1)} T^{(0)} $}

Finally this S-matrix element of  a  symmetric RR field ,one scalar  in zero picture and two tachyons
on the world volume of brane anti brane system  can be written down by
\beqa
 {\cal A}^ {C^{(-1)}  \phi^{(0)} T^{(-1)} T^{(0)}}& \sim & \int dx_1 dx_2 dx_3 d^2z \lan V_{\phi}^{(0)}{(x_{1})}
V_{T}^{(-1)}{(x_{2})}V_T^{(0)}{(x_{3})}
V_{RR}^{(-\frac{1}{2},-\frac{1}{2})}(z,\bar{z})\ran \nonumber\eeqa

with the following vertices 

\beqa
V_{T}^{(-1)}(y) &=&  e^{-\phi(y)} e^{2ik\cd X(y)}
\labell{vertex12}\otimes \sigma_2\\
V_{RR}^{(-\frac{1}{2},-\frac{1}{2})}(z,\bar{z})&=&(P_{-}\fsH_{(n)}M_p)^{\al\be}e^{-\phi(z)/2}
S_{\al}(z)e^{ip\cd X(z)}e^{-\phi(\bar{z})/2} S_{\be}(\bar{z})
e^{ip\cd D \cd X(\bar{z})}\otimes \sigma_3,\nonumber\eeqa

we just write down the amplitude in its compact form as

\beqa {\cal A}^{C^{(-1)}  \phi^{(0)}  T^{(-1)} T^{(0)}}&\sim& \int
 dx_{1}dx_{2}dx_{3}dx_{4} dx_{5}\,
(P_{-}\fsH_{(n)}M_p)^{\al\be}\xi_{1i}( 4 k_{3b}) x_{45}^{-1/4}(x_{24}x_{25})^{-1/2}I\nonumber\\&&\times\bigg((ip^i \frac{x_{54}}{x_{15}x_{14}})I_1^{b}+2i k_{1a}I_2^{bia}\bigg)\nonumber\eeqa where
$x_{ij}=x_i-x_j$,
\beqa
I&=&|x_{12}|^{4k_1.k_2}|x_{13}|^{4k_1.k_3}|x_{14}x_{15}|^{2k_1.p}|x_{23}|^{4k_2.k_3}|x_{24}x_{25}|^{2k_2.p}
|x_{34}x_{35}|^{2k_3.p}|x_{45}|^{p.D.p}\nonumber
\eeqa

where the following correlators need to be replaced inside the amplitude
\beqa
I_1^{b}&=&<:S_{\al}(x_4):S_{\be}(x_5):\psi^b(x_3):>
=2^{-1/2}x_{45}^{-3/4} (x_{34}x_{35})^{-1/2}(\gamma^{b}C^{-1})_{\alpha\beta}
\nonumber
\label{61re}\eeqa

  also

\beqa
I_2^{bia}&=&<:S_{\al}(x_4):S_{\be}(x_5):\psi^a \psi^i(x_1):\psi^b(x_3):>
=2^{-3/2}x_{45}^{1/4} (x_{34}x_{35})^{-1/2} (x_{14}x_{15})^{-1}\nonumber\\&&\bigg\{(\Gamma^{bia}C^{-1})_{\alpha\beta}+2\eta^{ab}\frac{Re[x_{14}x_{35}]}{x_{13}x_{45}}(\gamma^{i}C^{-1})_{\alpha\beta}\bigg\}
\nonumber
\label{6mre}\eeqa

Notice that, given the above correlations , one may easily investigate the
SL(2,R) transformation of the S-matrix. We performed  gauge fixing by just fixing the location of open strings so that the final integration needs to be done over the closed string RR 's position on the upper half complex plane. \footnote{ \beqar
 x_{1}=0 ,\qquad x_{2}=1,\qquad x_{3}\rightarrow \infty,
 \qquad dx_1dx_2dx_3\rightarrow x_3^{2}.
 \eeqar}

 Once more after gauge fixing one finds the same sort of the integration as we discussed earlier on. \footnote{
\beqa
 \int d^2 \!z |1-z|^{a} |z|^{b} (z - \bar{z})^{c}
(z + \bar{z})^{d},
 \eeqa
  }



\vskip.2in

Having calculated all the desired integrals, one could explore the result for the S-matrix in its particular picture  as follows

\beqa {\cal A}^{C^{(-1)}  \phi^{(0)}  T^{(-1)} T^{(0)}}&=&{\cal A}_{1}+{\cal A}_{2}+{\cal A}_{3}\labell{181u}\eeqa

in such a way that all its components are given by

\beqa
{\cal A}_{1}&\!\!\!\sim\!\!\!&-2i 2^{1/2} p^i\xi_{1i}k_{3b}
\Tr(P_{-}\fsH_{(n)}M_p\gamma^{b})L_1,
\nonumber\\
{\cal A}_{2}&\sim&-2i 2^{1/2}\xi_{1i} \bigg\{\Tr(P_{-}\fsH_{(n)}M_p \Gamma^{bia})\bigg\}k_{1a} k_{3b} L_1
\nonumber\\
{\cal A}_{3}&\sim&-2i 2^{1/2}\xi_{1i} \bigg\{\Tr(P_{-}\fsH_{(n)}M_p \gamma^{i})\bigg\} L_2
\labell{4k8}\eeqa
where
 $L_1,L_2$ functions are already given in \reef{kk43}. If we compare \reef{4k8} with \reef{4859}, then one might wonder whether the first term of ${\cal A}^{C^{(-1)}  \phi^{(0)}  T^{(-1)} T^{(0)}}$ is extra term , because it is also related to all infinite singularities. However, it is worth pointing that we have already produced all the infinite u-channel guage poles by taking into account the second term of \reef{4k8}. Therefore one has to find out or generalise new   Bianchi identities to actually  remove the first extra term of \reef{4k8}. Hence we apply momentum conservation $k_{1a}=-k_{2a}-k_{3a}-p_{a}$ to the second term of  \reef{4k8}, extract the traces, use the antisymmetric property of $\eps$ tensor  and eventually  add the first and second components of the amplitude to derive the following Bianchi identity

\beqa
\xi_{1i} k_{3b}\bigg(-p_a\eps^{a_{0}\cdots a_{p-2}ab}H^{i}_{a_{0}\cdots a_{p-2}}+p^i\eps^{a_{0}\cdots a_{p-1}b}H_{a_{0}\cdots a_{p-1}}\bigg)=0 \label{oo}\eeqa

Thus by holding \reef{oo} and keeping in mind momentum conservation along the world volume of brane we are precisely able to obtain the same  S-matrix as appeared in \reef{4859}, so we come to the crucial fact about this five point function of brane anti brane system  as follows.

\vskip.2in

 If we consider closed string RR in the asymmetric picture $C^{-3/2,-1/2}$ and scalar field in zero picture with some other open string tachyons then one must  find out all new Bianchi identities to be able to remove all the extra apparent singularities of asymmetric picture of brane/ anti brane systems.
\vskip.2in

 In the next section we will reveal some remarks for non-BPS branes   to  actually restore Ward identity associated (gauge invariance) to the gauge field and also we derive all the precise contact interactions of mixed RR, scalar/ gauge and tachyon string amplitudes. Namely we show that  if we consider the scalar field in zero picture and the RR in  asymmetric picture $(C^{-3/2,-1/2})$ then in this particular case there is no need to explore new Bianchi identities to the S-matrix of non-BPS amplitudes and more importantly those amplitudes independently  respect the Ward identity associated to the gauge field.




\vskip.5in


\section{ $C^{-1}\phi^{0}A^{-1}T^{0}$ S-matrix }

 In \cite{Hatefi:2013yxa} the five point world-sheet amplitude of a symmetric closed string RR with one scalar, a gauge field and a tachyon in the world volume of non-BPS branes of type II string theory 
 $<C^{-1}\phi^{-1}A^{0}T^{0}>$ was achieved and all the correlators were found.\footnote{

\beqa
{\cal A}^{C^{-1}\phi^{-1}A^{0}T^{0}} & \sim & \int dx_{1}dx_{2}dx_{3}dzd\bar{z}\,
  \lan V_{\phi}^{(-1)}{(x_{1})}
V_{A}^{(0)}{(x_{2})}V_T^{(0)}{(x_{3})}
V_{RR}^{(-\frac{1}{2},-\frac{1}{2})}(z,\bar{z})\ran,\labell{cor10}\eeqa

with  the following vertex operators :

\beqa
V_{\phi}^{(-1)}(y) &=& \xi_{1i}\psi^i(y)e^{\alpha'ik.X(y)}e^{-\phi(y)}\otimes \sigma_3\nonumber\\
V_{A}^{(0)}(x) &=& \xi_{2a}(\partial^a X(x)+i\alpha'q.\psi\psi^a(x))e^{\alpha'iq.X(x)}\otimes I
\nonumber
\eeqa
and 
 $u'=(-u-\frac{1}{4})$ and also
\beqa
L_1'&=&(2)^{-2(t+s+u)}\pi{\frac{\Gamma(-u+\frac{1}{4})
\Gamma(-s+\frac{1}{4})\Gamma(-t+\frac{1}{2})\Gamma(-t-s-u+\frac{1}{2})}
{\Gamma(-u-t+\frac{3}{4})\Gamma(-t-s+\frac{3}{4})\Gamma(-s-u+\frac{1}{2})}},\nonumber\\
L_3'&=&(2)^{-2(t+s+u)-1}\pi{\frac{\Gamma(-u-\frac{1}{4})
\Gamma(-s+\frac{3}{4})\Gamma(-t)\Gamma(-t-s-u)}
{\Gamma(-u-t+\frac{3}{4})\Gamma(-t-s+\frac{3}{4})\Gamma(-s-u+\frac{1}{2})}}
\nonumber\eeqa }

The final result in this symmetric picture was read off as

 \beqa {\cal A}^{C^{-1}\phi^{-1}A^{0}T^{0}}&=&{\cal A}_{1}+{\cal A}_{2}\labell{17u}\eeqa
where
\beqa
{\cal A}_{1}&\!\!\!\sim\!\!\!&2\xi_{1i}\xi_{2a}k_{3c}k_{2d}
\Tr(P_{-}\fsH_{(n)}M_p\Gamma^{cadi}
)L_1',
\nonumber\\
{\cal A}_{2}&\sim& (2L_3')\bigg\{-t\Tr(P_{-}\fsH_{(n)}M_p \gamma.\xi_2\gamma.\xi_{1})u'-2tk_3.\xi_2\Tr(P_{-}\fsH_{(n)}M_p \gamma.k_2\gamma.\xi_{1})
\nonumber\\&&
+\Tr(P_{-}\fsH_{(n)}M_p \gamma.k_3\gamma.\xi_{1})\bigg(-2t(k_3.\xi_2)+2u'k_1.\xi_2\bigg)\bigg\}
\labell{qqyt}\eeqa

It was also shown that one needs to use the momentum conservation as
$s+t+u=-p^ap_a-\frac{1}{4}$ ,  applying
$ t \rightarrow 0,  \quad  s \rightarrow  -\frac{1}{4} , \quad u \rightarrow  -\frac{1}{4} $ to the S-matrix to be able to derive all infinite $u',t$ channel tachyon and scalar poles of the non super symmetric amplitudes accordingly. Note that $L_1'$ has  just infinite contact interactions.

\vskip.2in

It is of high importance to note that this particular
${\cal A}^{C^{-1}\phi^{-1}A^{0}T^{0}}$ amplitude does respect all the symmetries and most importantly it does satisfy Ward identity  associated to the gauge field. Indeed if we replace  $\xi_{2a}\rightarrow k_{2a}$ inside \reef{qqyt} then the first term of \reef{qqyt} is automatically zero because

\beqa k_{2a} k_{2d} k_{3c} \eps^{a_{0}\cdots a_{p-2}cad}=0\nonumber\eeqa

  replacing $\xi_{2a}\rightarrow k_{2a}$ inside the second, third , fourth and fifth term of \reef{qqyt}  we also get zero result for the whole S-matrix \footnote{
\beqa
k_{2a} \xi_{1i} (u'-u')\Tr(P_{-}\fsH_{(n)}M_p \gamma^a\gamma^i)=0\nonumber\eeqa
 \beqa
 k_{3a} \xi_{1i} (tu'-tu')\Tr(P_{-}\fsH_{(n)}M_p \gamma^a\gamma^i)=0\nonumber\eeqa }
so gauge invariance is satisfied. Thus  based on just replacing $\xi_{2a}\rightarrow k_{2a}$ of the gauge field, the whole S-matrix vanishes.

\vskip.3in

 Having explained all the needed ingredients of the S-matrices , in the following  we would like to change the vertex of scalar field and see what happens to the gauge invariance of a mixture of five point world-sheet amplitude of a symmetric closed string RR with one scalar in zero picture , a gauge field and a tachyon in the world volume of non-BPS branes of type II string theory . This  $<C^{-1}\phi^{0}A^{-1}T^{0}>$ amplitude  is  given by  the following correlation functions

\beqa
{\cal A}^{C^{-1}\phi^{0}A^{-1}T^{0}} & \sim & \int dx_{1}dx_{2}dx_{3}dzd\bar{z}\,
  \lan V_{\phi}^{(0)}{(x_{1})}
V_{A}^{(-1)}{(x_{2})}V_T^{(0)}{(x_{3})}
V_{RR}^{(-\frac{1}{2},-\frac{1}{2})}(z,\bar{z})\ran,\labell{pp}\eeqa

Let us write down  the rest of the vertex operators including their CP factors as 

\beqa
V_{A}^{(-1)}(y) &=& \xi_{2a}\psi^a(y)e^{\alpha'iq.X(y)}e^{-\phi(y)}\otimes \sigma_3\nonumber\\
V_{\phi}^{(0)}(x) &=& \xi_{1i}(\partial^i X(x)+i\alpha'k.\psi\psi^i(x))e^{\alpha'ik.X(x)}\otimes I\nonumber\\
\eeqa

  where the following on-shell conditions for  scalar, gauge, RR and tachyon hold.\footnote{\beqa
 k^2=q^2=p^2=0, \quad  k_{3}^2=1/4  ,\quad q.\xi_1=k_2.\xi_1=0
\nonumber\eeqa} Once more we deal with just the holomorphic elements of  all fields involving   $X^{\mu}\psi^\mu, \phi$, so that the S-matrix is now given by

\beqa {\cal A}^{C^{-1}\phi^{0}A^{-1}T^{0}}&\sim& \int
 dx_{1}dx_{2}dx_{3}dx_{4} dx_{5}\,
(P_{-}\fsH_{(n)}M_p)^{\al\be}\xi_{1i}\xi_{2a}(\alpha'ik_{3c})x_{45}^{-1/4}(x_{24}x_{25})^{-1/2}\nonumber\\&&
\times(I_1+I_2)I\Tr(\lam_1\lam_2\lam_3)\Tr(\sig_3\sig_1I\sig_3\sig_1),\labell{125y}\eeqa 
where\footnote{
\beqa
I&=&|x_{12}|^{\alpha'^2k_1.k_2}|x_{13}|^{\alpha'^2k_1.k_3}|x_{14}x_{15}|^{\frac{\alpha'^2}{2}k_1.p}|x_{23}|^{\alpha'^2k_2.k_3}
|x_{24}x_{25}|^{\frac{\alpha'^2}{2}k_2.p}
|x_{34}x_{35}|^{\frac{\alpha'^2}{2} k_3.p}|x_{45}|^{\frac{\alpha'^2}{4}p.D.p}\eeqa}  $x_{ij}=x_i-x_j$.  Let us find out all the fermionic and bosonic correlators as
\beqa
I_1&=&{<:\partial X^i{(x_1)}e^{\alpha'ik_1.X(x_1)}: e^{\alpha'ik_2.X(x_2)}
 :e^{\alpha'ik_3.X(x_3)}:e^{\frac{\alpha'}{2}ip.X(x_4)}:e^{\frac{\alpha'}{2}ip.D.X(x_5)}:>}\nonumber \\&&\times{<:S_{\al}(x_4):S_{\be}(x_5):\psi^a(x_2):\psi^c(x_3)>},\nonumber\\
I_2&=&{<:e^{\alpha'ik_1.X(x_1)}: e^{\alpha'ik_2.X(x_2)}
 :e^{\alpha'ik_3.X(x_3)}:e^{\frac{\alpha'}{2}ip.X(x_4)}:e^{\frac{\alpha'}{2}ip.D.X(x_5)}:>}\nonumber \\&&
\alpha'ik_{1d} {<:S_{\al}(x_4):S_{\be}(x_5):\psi^d\psi^{i}(x_1):\psi^a(x_2)::\psi^c(x_3):>}
\eeqa

 We need to use the Wick-like rule~\cite{Liu:2001qa} to get to all the generalizations of the correlation functions of two spin and two fermion operators such as the following

\beqa
I_5^{ca}&=&<:S_{\al}(x_4):S_{\be}(x_5):\psi^a(x_2):\psi^c(x_3):>
=2^{-1}x_{45}^{-1/4} (x_{24}x_{25}x_{34}x_{35})^{-1/2}\nonumber\\&&\times\bigg\{(\Gamma^{ca}C^{-1})_{\alpha\beta}-2\frac{Re[x_{24}x_{35}]}{x_{23}x_{45}}\eta^{ac}(C^{-1})_{{\alpha\beta}}\bigg\}
\nonumber
\label{691}\eeqa

Now we need to make use of \cite{Hatefi:2010ik} to come over the final answer of the correlation  function in ten dimensions,

\beqa
I_6^{caid}&=&<:S_{\al}(x_4):S_{\be}(x_5):\psi^d\psi^{i}(x_1):\psi^a(x_2)::\psi^c(x_3):>\nonumber\\&&
=\bigg\{(\Gamma^{caid}C^{-1})_{{\alpha\beta}}+\frac{Re[x_{14}x_{25}]}{x_{12}x_{45}}(2\eta^{da}(\Gamma^{ci}C^{-1})_{{\alpha\beta}})-2\frac{Re[x_{14}x_{35}]}{x_{13}x_{45}}\eta^{dc}(\Gamma^{ai}C^{-1})_{{\alpha\beta}}\nonumber\\&&-2\frac{Re[x_{24}x_{35}]}{x_{23}x_{45}}\eta^{ac}(\Gamma^{id}C^{-1})_{{\alpha\beta}}\bigg\}
2^{-2}x_{45}^{3/4}(x_{24}x_{25}x_{34}x_{35})^{-1/2}(x_{14}x_{15})^{-1},\label{hh}\eeqa

Now we are allowed to replace all the correlators inside \reef{125y} to obtain the compact form of the desired S-matrix as follows

\beqa
{\cal A}^{C^{-1}\phi^{0}A^{-1}T^{0}}&\!\!\!\!\sim\!\!\!\!\!&\int dx_{1}dx_{2} dx_{3}dx_{4}dx_{5}(P_{-}\fsH_{(n)}M_p)^{\al\be}I\xi_{1i}\xi_{2a}(-2\alpha'ik_{3c}) x_{45}^{-1/4}(x_{24}x_{25})^{-1/2}\nonumber\\&&\times
\bigg(a^i_2 I_5^{ca} +\alpha'ik_{1d}I_6^{caid}\bigg)\Tr(\lam_1\lam_2\lam_3)\labell{amp3},\eeqa
so that
\beqa
a^i_2&=&ip^{i}\bigg(\frac{x_{54}}{x_{14}x_{15}}\bigg)
\eeqa

We are now able to investigate that the amplitude has satisfied the $SL(2,R)$ invariance, we also did gauge fixing by just fixing the positions of three open strings.\footnote{
 \beqa
 x_{1}=0 ,\qquad x_{2}=1,\qquad x_{3}\rightarrow \infty,
 \eeqa
 we lead to \beqa
 \int d^2 \!z |1-z|^{a} |z|^{b} (z - \bar{z})^{c}
(z + \bar{z})^{d},
 \eeqa with following Mandelstam variables
\beqar
s&=&-\frac{\alpha'}{2}(k_1+k_3)^2,\qquad t=-\frac{\alpha'}{2}(k_1+k_2)^2,\qquad u=-\frac{\alpha'}{2}(k_2+k_3)^2.
\qquad\eeqar see \cite{Fotopoulos:2001pt}}

\vskip.3in

The solutions for all the integrals on upper half plane have been released and  the  ultimate result of the S-matrix will be obtained as
\beqa {\cal A}^{C^{-1}\phi^{0}A^{-1}T^{0}}&=&{\cal A}_{1}+{\cal A}_{2},\labell{11u}\eeqa

where
\beqa
{\cal A}_{1}&\!\!\!\sim\!\!\!&\bigg(2\xi_{1i}\xi_{2a}k_{3c}k_{1d}
\Tr(P_{-}\fsH_{(n)}M_p\Gamma^{caid})-\xi_1.p(2k_{3c}\xi_{2a}) \Tr(P_{-}\fsH_{(n)}M_p\Gamma^{ca})
\bigg)L_1',
\nonumber\\
{\cal A}_{2}&\sim&  \bigg\{t\xi_1.p (4k_{3}.\xi_2)\Tr(P_{-}\fsH_{(n)}M_p )+ 4(u+\frac{1}{4})k_{3c}\xi_{1i}\Tr(P_{-}\fsH_{(n)}M_p\Gamma^{ci})k_1.\xi_2\nonumber\\&&-4tk_{3}.\xi_{2} k_{1b}\xi_{1i}\Tr(P_{-}\fsH_{(n)}M_p\Gamma^{ib})-2t(u+\frac{1}{4}) \xi_{1i}\xi_{2a}\Tr(P_{-}\fsH_{(n)}M_p\Gamma^{ai})
\bigg\}L_3'
\labell{497m}\eeqa

Note that we have already analysed all infinite $u'$ tachyon and massless $t$  channel scalar poles of the amplitude in 
\cite{Hatefi:2013yxa}. 
\vskip.2in

In this picture after replacing   $\xi_{2a}\rightarrow k_{2a}$ ( due to the terms $\xi_1.p$ ) the amplitude does not  satisfy Ward identity associated to the gauge field and indeed the second and third terms would remain whereas $L_3'$ can not be removed by $L_1'$. Let us compare the result of this amplitude \reef{11u} with \reef{qqyt}  to make a statement on mixed closed and open string amplitudes  including one scalar and one RR and some other open tachyons. The last term in \reef{497m} is exactly the second term of \reef{qqyt}, the fourth term in \reef{497m} is exactly the last term of \reef{qqyt}. Note  that if we add the third and fourth term of  \reef{qqyt} and use the momentum conservation along the world volume of brane , then the result is precisely equivalent with the fifth term of \reef{497m}. Once more by using momentum conservation in world volume direction
 the first term of \reef{497m} is exactly equivalent  with the first term in \reef{qqyt}.

 \vskip.2in

 However, the second and the third terms of \reef{497m} are extra terms and in particular if we use anti commutator relation of  $\gamma$ matrices these two terms can not cancel each other due to the fact that $L_3'$ is different from $L_1'$ .

  \vskip.2in

 Indeed if we replace  $\xi_{2a}\rightarrow k_{2a}$ inside \reef{497m} and use momentum conservation along the brane , the first term is automatically zero because
\beqa k_{2a} k_{3c} k_{1d} \eps^{a_{0}\cdots a_{p-2}cad}=0\nonumber\eeqa

 By replacing $\xi_{2a}\rightarrow k_{2a}$ inside the  fourth, fifth and sixth  terms of \reef{497m}  appropriately  we also get zero result as follows
\beqa
-2tu' \xi_{1i} (k_{2a}+ k_{1a}+k_{3a})\eps^{a_{0}\cdots a_{p-1}a}=0\nonumber\eeqa
 Thus the second and third terms  give rise the amplitude not to be gauge invariant unless one finds out some  new Bianchi identities.  \footnote{The resolution for this problem (to get satisfied gauge invariance of the above S-matrix) is to add up the third term of  \reef{497m} with the the other terms in ${\cal A}_{2}$ of \reef{497m} and also do add the first and second term of \reef{497m} together to actually get to the so called new identities as follows

  \beqa
  \xi_{1i} (p^i \eps^{a_{0}\cdots a_{p}}H_{a_{0}\cdots a_{p}}-p_c\eps^{a_{0}\cdots a_{p-1}c}H^{i}_{a_{0}\cdots a_{p-1}})=0\nonumber\\
  \xi_{1i} k_{3c} k_{2a}(-p_d\eps^{a_{0}\cdots a_{p-3}cad}H^{i}_{a_{0}\cdots a_{p-3}}+p^i \eps^{a_{0}\cdots a_{p-2}ca}H_{a_{0}\cdots a_{p-2}})=0\nonumber\\
  \eeqa. }

 In the next section we show that by considering the asymmetric  RR and a scalar, a gauge  and one tachyon in zero picture of non-BPS branes, the S-matrix automatically satisfies Ward identity without needing to introduce any new Bianchi identities.

\subsection{ $C^{-2}\phi^{0}A^{0}T^{0}$ S-matrix}

One can  do the  same CFT methods  to actually derive  the entire S-matrix of the above strings in the asymmetric picture. Hence the final answer for   the five point world-sheet amplitude of a closed string RR (in asymmetric picture) with one scalar, a gauge field and a tachyon in the world volume of non-BPS branes of type II string theory  $<C^{-2}\phi^{0}A^{0}T^{0}>$ is

 \beqa {\cal A}^{C^{-2}\phi^{0}A^{0}T^{0}}&=&{\cal A}_{1}+{\cal A}_{2}+{\cal A}_{3}+{\cal A}_{4}\labell{1u8u}\eeqa
where
\beqa
{\cal A}_{1}&\sim & 2^{3/2}i\xi_{1i}\xi_{2a}k_{3c}k_{2b}L_1'
\bigg(p^i\Tr(P_{-}\fsC_{(n)}M_p\Gamma^{cab})-k_{1d}\Tr(P_{-}\fsC_{(n)}M_p\Gamma^{cabid}\bigg)
\nonumber\\
{\cal A}_{2}&\sim&2^{3/2}i\xi_1.p L_3' \Tr(P_{-}\fsC_{(n)}M_p \gamma^c)\bigg(2tk_3.\xi_2  [-k_{3c}-k_{2c}]
+2k_1.\xi_2u'k_{3c}-tu'\xi_{2c}\bigg)
\nonumber\\
{\cal A}_{3}&\sim &2^{3/2}i\xi_{1i}L_3' \Tr(P_{-}\fsC_{(n)}M_p \Gamma^{cid})\bigg[-2k_1.\xi_2 u'k_{3c}(k_{1d}+k_{2d})+2tk_3.\xi_2 k_{1d}(k_{3c}+k_{2c})\bigg]
\nonumber\\
{\cal A}_{4}&\sim &2^{3/2}i\xi_{1i}L_3't u' \xi_{2a} \Tr(P_{-}\fsC_{(n)}M_p \Gamma^{cai})(k_{3c}+k_{1c}+k_{2c})
\labell{qq689n}\eeqa

 with
 $u'=(-u-\frac{1}{4})$ and the same introduced $L_1',L_3'$ . $L_1'$ has infinite contact interactions and $L_3'$ has infinite $t,u'$ scalar-tachyon channel poles accordingly.

\vskip.2in

The nice thing about this asymmetric S-matrix is that without introducing any further Bianchi identity this amplitude automatically does satisfy the Ward identity associated to the gauge field  which means that if we replace $\xi_{2a}\rightarrow k_{2a}$ the whole S-matrix vanishes where the following points are needed.
 In the first term of ${\cal A}_{2}$ one has to apply momentum conservation along the world volume of brane $-k_{3c}-k_{2c}=p_c+k_{1c}$ and apply the following identity

 \beqa
 p_c \eps^{a_{0}...a_{p-1}c}&=&0.
 \nonumber\eeqa

 We need to apply  momentum conservation in ${\cal A}_{3}$'s  first term, that is  $-k_{1d}-k_{2d}=p_d+k_{3d}$ and then  draw particular attention to the fact that this part of the S-matrix  involves $k_{3d} k_{3c}\eps^{a_{0}...a_{p-2}dc} C^{i}_{a_{0}...a_{p-2}}$  which is zero because of the antisymmetric property of the $\eps$ tensor. Likewise, we need to replace $k_{3c}+k_{2c}=-p_c-k_{1c}$ and $k_{1d} k_{1c}\eps^{a_{0}...a_{p-2}dc} C^{i}_{a_{0}...a_{p-2}}=0$. Finally we apply momentum conservation to the last term of \reef{qq689n} and note to the point that $p_c \eps^{a_{0}...a_{p-2}ca} C^{i}_{a_{0}...a_{p-2}}=0$ plays the crucial role in checking the Ward identity.

 \vskip.2in

 The last remark about the asymmetric picture of the S-matrices is that , one finds out all the entire contact interactions of string theory amplitudes, properly. As an instance  this amplitude  includes several contact interaction  terms like the first term and the last terms of \reef{qq689n}  which could be missed in its symmetric picture of     \reef{497m}.

 \section{Conclusion}

We have derived scattering amplitudes of all  three , four and five point BPS and non-BPS mixture of a closed string Ramond-Ramond , a scalar field , gauge and tachyons in  their all different pictures of both world volume and transverse directions (for general $p,n$ cases) of  type IIA (IIB) String theory.
\vskip.1in

In particular we have shown that if we carry out the calculations of higher point functions in asymmetric picture of Ramond-Ramond (taking its vertex operator in terms of its potential $C^{-2}$) and  scalar field
in zero picture then, various equations must be kept fixed for BPS branes, the entire contact interactions can be definitely obtained and most importantly the Ward identity associated to the gauge field is  also satisfied.

More accurately, by direct calculations on upper half plane, we have also observed   that  some of the Bianchi identities (that must be true) for BPS branes can not be necessarily applied to non-BPS  amplitudes, otherwise the whole S-matrix might be vanished. Indeed  in the presence of the scalar field and RR, the terms carrying momentum of RR in transverse directions $(p^i,p^j)$ play important role in the entire form of the S-matrix and one has to keep them  in five point functions.

 \vskip.1in

 We expect it to be true for higher point functions of string theory amplitudes and it would be nice to check it directly.  It would also be important to deal with some  other subtleties of the perturbative string theory \cite{Witten:2012bh}.


\section*{Acknowledgments}

I would like to thank E.Witten, J. Polchinski, L.Alvarez-Gaume, W.Lerche, R. de Mello Koch , S.Ramgoolam, R.Myers, S.Hirano, V.Jejjala, K.Zoubos, D.Friedan , N.Arkani-Hamed and K.S. Narain  for valuable discussions/comments and for their great remarks. Some parts of the  computations of this paper were done in Harvard, Institute for advanced study in Princeton, ICTP and CERN, but the final work has been completed in IHES. In  particular, the author  would like to thank both ICTP and CERN for their great hospitality and  their continuous  supports in his career.


\begin{thebibliography}{2007}

\bibitem{Polchinski:1995mt}
  J.~Polchinski,
  ``Dirichlet Branes and Ramond-Ramond charges,''
  Phys.\ Rev.\ Lett.\  {\bf 75}, 4724 (1995)
  [hep-th/9510017].

\bibitem{Witten:1995im}
  E.~Witten,
  ``Bound states of strings and p-branes,''
  Nucl.\ Phys.\ B {\bf 460}, 335 (1996)
  [hep-th/9510135].
\bibitem{Polchinski:1996na}
  J.~Polchinski,
  ``Tasi lectures on D-branes,''
  hep-th/9611050.

\bibitem{Myers:1999ps}
  R.~C.~Myers,
  ``Dielectric branes,''
  JHEP {\bf 9912} (1999) 022
  [hep-th/9910053].
\bibitem{Hatefi:2012sy}
  E.~Hatefi, A.~J.~Nurmagambetov and I.~Y.~Park,
  ``$N^3$ entropy of $M5$ branes from dielectric effect,''
  Nucl.\ Phys.\ B {\bf 866}, 58 (2013)
  [arXiv:1204.2711 [hep-th]].




\bibitem{Bachas:1995kx}
  C.~Bachas,
  ``D-brane dynamics,''
  Phys.\ Lett.\ B {\bf 374}, 37 (1996)
  [hep-th/9511043].
\bibitem{Li:1995pq}
  M.~Li,
  ``Boundary states of D-branes and Dy strings,''
  Nucl.\ Phys.\ B {\bf 460}, 351 (1996)
  [hep-th/9510161].
\bibitem{Douglas:1995bn}
  M.~R.~Douglas,
  ``Branes within branes,''
  In *Cargese 1997, Strings, branes and dualities* 267-275
  [hep-th/9512077].
\bibitem{Green:1996dd}
  M.~B.~Green, J.~A.~Harvey and G.~W.~Moore,
  ``I-brane inflow and anomalous couplings on d-branes,''
  Class.\ Quant.\ Grav.\  {\bf 14} (1997) 47
  [hep-th/9605033].

\bibitem{Junghans:2014zla}
  D.~Junghans and G.~Shiu,
   ``Brane Curvature Corrections to the $\mathcal{N}=1$ Type II/F-theory Effective Action,''
  arXiv:1407.0019 [hep-th].


\bibitem{Hatefi:2012zh}
  E.~Hatefi,
   ``Shedding light on new Wess-Zumino couplings with their corrections to all orders in alpha-prime,''
  JHEP {\bf 1304}, 070 (2013)
  [arXiv:1211.2413 [hep-th]].




\bibitem{Hatefi:2010ik}
  E.~Hatefi,
  ``On effective actions of BPS branes and their higher derivative corrections,''
  JHEP {\bf 1005}, 080 (2010)
  [arXiv:1003.0314 [hep-th]].


\bibitem{Hatefi:2012wj}
  E.~Hatefi,
   ``On higher derivative corrections to Wess-Zumino and Tachyonic actions in type II super string theory,''
  Phys.\ Rev.\ D {\bf 86}, 046003 (2012)
  [arXiv:1203.1329 [hep-th]];
  E.~Hatefi,
  ``All order $\alpha'$ higher derivative corrections to non-BPS branes of type IIB Super string theory,''
  JHEP {\bf 1307}, 002 (2013)
  [arXiv:1304.3711 [hep-th]].

\bibitem{Sen:1999mg}
  A.~Sen,
  ``NonBPS states and Branes in string theory,''
  hep-th/9904207.
\bibitem{Lambert:2003zr}
  N.~D.~Lambert, H.~Liu and J.~M.~Maldacena,
   ``Closed strings from decaying D-branes,''
  JHEP {\bf 0703}, 014 (2007)
  [hep-th/0303139].
\bibitem{Sen:2004nf}
  A.~Sen,
   ``Tachyon dynamics in open string theory,''
  Int.\ J.\ Mod.\ Phys.\ A {\bf 20}, 5513 (2005)
  [hep-th/0410103].
\bibitem{Hatefi:2013eia}
  E.~Hatefi,
  ``Closed string Ramond-Ramond proposed higher derivative interactions on fermionic amplitudes in IIB,''
  Nucl.\ Phys.\ B {\bf 880}, 1 (2014)
  [arXiv:1302.5024 [hep-th]];
  E.~Hatefi,
  ``Super-Yang-Mills, Chern-Simons couplings and their all order $\alpha '$ corrections in IIB superstring theory,''
  Eur.\ Phys.\ J.\ C {\bf 74}, no. 8, 3003 (2014)
  [arXiv:1310.8308 [hep-th]];
  E.~Hatefi,
  ``SYM, Chern-Simons, Wess-Zumino Couplings and their higher derivative corrections in IIA Superstring theory,''
  Eur.\ Phys.\ J.\ C {\bf 74}, 2949 (2014)
  [arXiv:1403.1238 [hep-th]];
  E.~Hatefi,
  ``More on Ramond-Ramond, SYM, WZ couplings and their corrections in IIA,''
  Eur.\ Phys.\ J.\ C {\bf 74}, no. 10, 3116 (2014)
  [arXiv:1403.7167 [hep-th]].

\bibitem{Barreiro:2013dpa}
  L.~A.~Barreiro and R.~Medina,
  ``RNS derivation of N-point disk amplitudes from the revisited S-matrix approach,''
  Nucl.\ Phys.\ B {\bf 886}, 870 (2014)
  [arXiv:1310.5942 [hep-th]];
  L.~A.~Barreiro and R.~Medina,
  ``Revisiting the S-matrix approach to the open superstring low energy effective lagrangian,''
  JHEP {\bf 1210}, 108 (2012)
  [arXiv:1208.6066 [hep-th]].

\bibitem{Hatefi:2012rx}
  E.~Hatefi and I.~Y.~Park,
   ``Universality in all-order $\alpha'$ corrections to BPS/non-BPS brane world volume theories,''
  Nucl.\ Phys.\ B {\bf 864}, 640 (2012)
  [arXiv:1205.5079 [hep-th]].








\bibitem{Maxfield:2013wka}
  T.~Maxfield, J.~McOrist, D.~Robbins and S.~Sethi,
  JHEP {\bf 1312}, 032 (2013)
  [arXiv:1309.2577 [hep-th]];
  E.~Hatefi, A.~J.~Nurmagambetov and I.~Y.~Park,
   ``Near-Extremal Black-Branes with n*3 Entropy Growth,''
  Int.\ J.\ Mod.\ Phys.\ A {\bf 27}, 1250182 (2012)
  [arXiv:1204.6303 [hep-th]].

\bibitem{Hatefi:2012bp}
  E.~Hatefi, A.~J.~Nurmagambetov and I.~Y.~Park,
  ``ADM reduction of IIB on $\mathcal{H}^{p,q}$ to dS braneworld,''
  JHEP {\bf 1304}, 170 (2013)
  [arXiv:1210.3825 [hep-th]].
\bibitem{deAlwis:2013gka}
  S.~de Alwis, R.~Gupta, E.~Hatefi and F.~Quevedo,
  ``Stability, Tunneling and Flux Changing de Sitter Transitions in the Large Volume String Scenario,''
  JHEP {\bf 1311}, 179 (2013)
  [arXiv:1308.1222 [hep-th], arXiv:1308.1222].




\bibitem{Bergman:1998xv}
  O.~Bergman and M.~R.~Gaberdiel,
   ``Stable nonBPS D particles,''
  Phys.\ Lett.\ B {\bf 441}, 133 (1998)
  [hep-th/9806155]
 ;
  M.~Frau, L.~Gallot, A.~Lerda and P.~Strigazzi,
  ``Stable nonBPS D-branes in type I string theory,''
  Nucl.\ Phys.\ B {\bf 564}, 60 (2000)
  [hep-th/9903123]
  ;
  E.~Dudas, J.~Mourad and A.~Sagnotti,
  ``Charged and uncharged D-branes in various string theories,''
  Nucl.\ Phys.\ B {\bf 620}, 109 (2002)
  [hep-th/0107081]
  ;
  E.~Eyras and S.~Panda,
   ``The Space-time life of a nonBPS D particle,''
  Nucl.\ Phys.\ B {\bf 584} (2000) 251
  [hep-th/0003033]
  ;
  E.~Eyras and S.~Panda,
  ``NonBPS branes in a type I orbifold,''
  JHEP {\bf 0105}, 056 (2001)
  [hep-th/0009224].

\bibitem{Witten:1998cd}
  E.~Witten,
  ``D-branes and K theory,''
  JHEP {\bf 9812}, 019 (1998)
  [hep-th/9810188].


\bibitem{Dvali:1998pa}
  G.~R.~Dvali and S.~H.~H.~Tye,
  ``Brane inflation,''
  Phys.\ Lett.\ B {\bf 450} (1999) 72
  [hep-ph/9812483].

\bibitem{Burgess:2001fx}
  C.~P.~Burgess, M.~Majumdar, D.~Nolte, F.~Quevedo, G.~Rajesh and R.~-J.~Zhang,
   ``The Inflationary brane anti-brane universe,''
  JHEP {\bf 0107}, 047 (2001)
  [hep-th/0105204].
\bibitem{Choudhury:2003vr}
  D.~Choudhury, D.~Ghoshal, D.~P.~Jatkar and S.~Panda,
   ``Hybrid inflation and brane - anti-brane system,''
  JCAP {\bf 0307}, 009 (2003)
  [hep-th/0305104].

\bibitem{Kachru:2003sx}
  S.~Kachru, R.~Kallosh, A.~D.~Linde, J.~M.~Maldacena, L.~P.~McAllister and S.~P.~Trivedi,
   ``Towards inflation in string theory,''
  JCAP {\bf 0310}, 013 (2003)
  [hep-th/0308055].

\bibitem{Casero:2007ae}
  R.~Casero, E.~Kiritsis and A.~Paredes,
   ``Chiral symmetry breaking as open string tachyon condensation,''
  Nucl.\ Phys.\ B {\bf 787}, 98 (2007)
  [hep-th/0702155 [HEP-TH]].
\bibitem{Dhar:2007bz}
  A.~Dhar and P.~Nag,
   ``Sakai-Sugimoto model, Tachyon Condensation and Chiral symmetry Breaking,''
  JHEP {\bf 0801}, 055 (2008)
  [arXiv:0708.3233 [hep-th]].


\bibitem{Friedan:1985ge}
  D.~Friedan, E.~J.~Martinec and S.~H.~Shenker,
  ``Conformal Invariance, Supersymmetry and String Theory,''
  Nucl.\ Phys.\ B {\bf 271} (1986) 93;






\bibitem{Kennedy:1999nn}
  C.~Kennedy and A.~Wilkins,
  ``Ramond-Ramond couplings on Brane - anti-Brane systems,''
  Phys.\ Lett.\ B {\bf 464}, 206 (1999)
  [hep-th/9905195].





\bibitem{Hatefi:2012ve}
  E.~Hatefi and I.~Y.~Park,
  ``More on closed string induced higher derivative interactions on D-branes,''
  Phys.\ Rev.\ D {\bf 85}, 125039 (2012)
  [arXiv:1203.5553 [hep-th]].



\bibitem{Witten:2012bh}
  E.~Witten,
  ``Superstring Perturbation Theory Revisited,''
  arXiv:1209.5461;
  E.~Witten,
  ``Notes On Super Riemann Surfaces And Their Moduli,''
  arXiv:1209.2459;
  E.~Witten,
  ``More On Superstring Perturbation Theory,''
  arXiv:1304.2832;
  E.~Witten,
  ``Notes On Supermanifolds and Integration,''
  arXiv:1209.2199;



\bibitem{Hatefi:2013yxa}
  E.~Hatefi,
  ``Selection Rules and RR Couplings on Non-BPS Branes,''
  JHEP {\bf 1311}, 204 (2013)
  [arXiv:1307.3520].



\bibitem{Bianchi:1991eu} 
  M.~Bianchi, G.~Pradisi and A.~Sagnotti,
  Nucl.\ Phys.\ B {\bf 376}, 365 (1992).



















\bibitem{Liu:2001qa}
  H.~Liu and J.~Michelson,
  `*-trek III: The search for Ramond-Ramond couplings,''
  Nucl.\ Phys.\  B {\bf 614}, 330 (2001)
  [arXiv:hep-th/0107172].

\bibitem{Garousi:2008ge}
  M.~R.~Garousi and E.~Hatefi,
 ``More on WZ action of non-BPS branes,''
  JHEP {\bf 0903}, 008 (2009)
  [arXiv:0812.4216 [hep-th]].


\bibitem{Hatefi:2011jq}
  E.~Hatefi,
   ``Three Point Tree Level Amplitude in Superstring Theory,''
  Nucl.\ Phys.\ Proc.\ Suppl.\  {\bf 216}, 234 (2011)
  [arXiv:1102.5042 [hep-th]].




\bibitem{Michel:2014lva}
  B.~Michel, E.~Mintun, J.~Polchinski, A.~Puhm and P.~Saad,
  ``Remarks on brane and antibrane dynamics,''
  arXiv:1412.5702 [hep-th].



\bibitem{Garousi:2007fk}
  M.~R.~Garousi and E.~Hatefi,
   ``On Wess-Zumino terms of Brane-Antibrane systems,''
  Nucl.\ Phys.\ B {\bf 800}, 502 (2008)
  [arXiv:0710.5875 [hep-th]].


\bibitem{Hatefi:2012cp}
  E.~Hatefi,
   ``On D-brane anti D-brane effective actions and their corrections to all orders in alpha-prime,''
  JCAP {\bf 1309}, 011 (2013)
  [arXiv:1211.5538, [hep-th]].

\bibitem{Fotopoulos:2001pt}
  A.~Fotopoulos,
   ``On (alpha')**2 corrections to the D-brane action for non-geodesic
  world-volume embeddings,''
  JHEP {\bf 0109}, 005 (2001)
  [arXiv:hep-th/0104146].
























\end{thebibliography}
\end{document}